\documentclass[fleqn,usenatbib]{mnras}

\usepackage{newtxtext,newtxmath}

\usepackage[T1]{fontenc}

\DeclareRobustCommand{\VAN}[3]{#2}
\let\VANthebibliography\thebibliography
\def\thebibliography{\DeclareRobustCommand{\VAN}[3]{##3}\VANthebibliography}

\usepackage{graphicx}	
\usepackage{amsmath}	
\usepackage{amssymb}	
\usepackage{caption}
\usepackage{subcaption}

\title[Relativistic oblique shocks revisited]{Relativistic oblique shocks with ordered or random magnetic fields: tangential field governs}

\author[J.-Z. Ma and B. Zhang]{
Jing-Ze Ma$^{1}$\thanks{E-mail: mjz18@mails.tsinghua.edu.cn}
and Bing Zhang$^{2}$\thanks{E-mail: zhang@physics.unlv.edu}
\\
$^{1}$School of Aerospace Engineering, Tsinghua University,
 Beijing 100084, China\\
$^{2}$Department of Physics and Astronomy, University of Nevada Las Vegas, Las Vegas, NV 89154, USA
}

\date{Accepted XXX. Received YYY; in original form ZZZ}

\pubyear{2021}

\begin{document}
\label{firstpage}
\pagerange{\pageref{firstpage}--\pageref{lastpage}}
\maketitle

\begin{abstract}
Relativistic magnetohydrodynamic shocks are efficient particle accelerators, often invoked in the models of gamma-ray bursts (GRBs) and shock-powered fast radio bursts (FRBs).
Most theoretical studies assume a perpendicular shock with an ordered magnetic field perpendicular to the shock normal.
However, the degree of magnetization $\sigma$ and the magnetic field geometry in shock-powered GRB/FRB scenarios are still poorly constrained by observations.
Analogous to the magnetization $\sigma$ associated with the total field strength, we define a tangential magnetization $\sigma_\perp$ associated with the tangential field component.
We explore the jump conditions of magnetized relativistic shocks, either with an ordered field of arbitrary inclination angle or with a random field of arbitrary anisotropy.
In either case, we find that the jump conditions of relativistic shocks are governed by the tangential magnetization $\sigma_\perp$ instead of the total magnetization $\sigma$, insensitive to the inclination angles or the anisotropy of the pre-shock magnetic field.
The approximated analytical solution developed in this work could serve as a quick check for numerical simulations and apply to theoretical studies of GRBs/FRBs with a more general field geometry.

\end{abstract}




\section{Introduction}

Relativistic shocks are often involved in astrophysical phenomena associated with compact objects.
They are believed to operate as efficient particle accelerators, producing nonthermal emission in pulsar wind nebulae \citep[e.g.][]{kennelConfinement1984A}, active galactic nuclei \citep[e.g.][]{blandfordRelativistic1979A} and gamma-ray bursts \citep[GRBs, e.g.][]{reesRelativistic1992M}. Recently, the shock-powered synchrotron maser emission is also invoked as an explanation for fast radio bursts \citep[FRBs, e.g.][]{lyubarskymodel2014M}, though the viability of relevant models is still under debate \citep{luradiation2018M, plotnikovsynchrotron2019M} and to be tested against observations \citep[for a recent review]{luoDiverse2020N, nimmoHighly2021NA, yuconfrontation2021M, libimodal2021N, zhangphysical2020N}.

When a relativistic ejecta collides with a surrounding medium (e.g., a previous ejecta, a magnetar wind, or an interstellar medium (ISM)), the standard shock-powered GRB/FRB scenarios invoke the formation of a pair of shocks if the conditions are appropriate, as indicated by the solution to the Riemann problem \citep{goedbloedAdvanced2010AM, 2013rehy.book.....R}.
The reverse shock propagates backward into the ejecta while the forward shock propagates into the medium.
Depending on the sites of collision, standard GRB theories predict the formation of internal shocks (faster part catching up with the slower part inside the outflow) as well as external shocks (the outflow colliding with the ISM) \citep{zhangPhysics2018TPoGBbBZI9CUP}.
The GRB prompt emission is attributed to the internal shocks or magnetic reconnections, while the afterglow is related to the external shocks.
The GRB-like models for FRBs have followed similar ideas.
It was proposed that a FRB could be produced by a flare colliding into the magnetar wind nebulae \citep{lyubarskymodel2014M}, into a previous flare \citep{metzgerFast2019M}, or into the magnetar wind itself \citep{beloborodovFlaring2017A, beloborodovBlast2020A}.

To accelerate particles to high energies, the relativistic shocks have to be permeated by  magnetic field lines (i.e., relativistic magnetohydrodynamic (MHD) shock), whether the fields are advected from the central engines or generated \textit{in situ}.
Within the context of GRB/FRB theories, the degree of magnetization of the fluid is measured by the parameter $\sigma$, the ratio of the magnetic field enthalpy density to the enthalpy density of matter in the fluid comoving frame (see equation~\eqref{eq:sigma} for a precise definition).
The magnetization of ISM is typically estimated as $\sigma_{\rm ISM} \sim 10^{-9}$ \citep{santanaMagnetic2014A}, whereas the initial magnetization of the ejecta near the central engines (e.g., magnetars) can be extremely high as $\sigma_0 \gg 1$.
The general theoretical framework of GRBs considers two types of jet models: a matter-dominated outflow \citep['fireball' driven by central engines with $\sigma_0 \ll 1$;][]{paczynskiGammaray1986A, goodmanAre1986A, shemiAppearance1990A} and a Poynting-flux-dominated outflow \citep[driven by central engines with $\sigma_0 \gg 1$;][]{usovNature1994M, thompsonmodel1994M, meszarosPoynting1997A, vlahakisRelativistic2003A, lyutikovGamma2003a}, although the magnetization of the latter case is expected to decrease as the jet propagates to a larger radius.

The magnetization of the ejecta at the initial launching site and the emission site are still poorly constrained by observations \citep{granotGammaRay2015SSR}, even though  some estimations on the upper limits of $\sigma$ have been made based on shock models and particle acceleration mechanisms \citep[e.g.][]{sironiRelativistic2015SSR}.
Observations of GRB afterglow light curves imply that the reverse shock is more magnetized than the forward shock at least for some GRBs, but still favor a reverse shock weakly or moderately magnetized \citep[$\sigma \lesssim 1$;][]{zhangGammaRay2003A, harrisonMagnetization2013A, japeljPhenomenology2014A, huangVery2016A, jordana-mitjansLowly2020A}.
For typical GRB conditions, shock modelling suggests that the external reverse shock likely does not exist when the upstream magnetization $\sigma$ exceeds a critical value defined by the bulk Lorentz factor of the jet and the density ratio between the two media (e.g. $\gtrsim 100$ for GRB problems) \citep{zhangGamma2005Aa, 2009ApJ...690L..47M, aimechanical2021M}, and the strength of internal shocks or 
external reverse shock would be suppressed if $\sigma \gtrsim 1$ \citep{mimicaDeceleration2009A, 2010MNRAS.401..525M, narayanConstraints2011M, sironiRelativistic2015M}.
In contrast, the shock-powered FRB models invoke more magnetized scenarios, with a $\sigma \gtrsim 1$ ejecta colliding into a $\sigma \gtrsim 10^{-3}$ medium \citep{metzgerFast2019M, beloborodovBlast2020A}.
Numerical simulations indicate that particle acceleration is only efficient if $\sigma \lesssim 10^{-3}$, where particles diffuse back and forth across the shock front and gain energy via diffusive shock acceleration \citep{sironiParticle2009A, sironiParticle2011A, sironiMaximum2013A, sironiRelativistic2015SSR}.
However, recent simulations show that the synchrotron maser instability would generate electromagnetic precursor waves propagating upstream from the shock front \citep{iwamotoPersistence2017A, iwamotoPrecursor2018A}, accelerating particles via wakefield acceleration with efficiency $10^{-3} \sigma^{-1}$ for $0.1 \lesssim \sigma \lesssim 10$ as measured in the downstream rest frame \citep{plotnikovsynchrotron2019M, sironiCoherent2021PRL}
\footnote{
This efficiency measures the fraction of shock energy converted into coherent radiation, which depends on the reference frame.
For shock-powered synchrotron maser emission in a pair plasma with $0.1 \lesssim \sigma \lesssim 10$, the efficiency was found to be $7\times 10^{-4} \sigma^{-2}$ in the shock rest frame and $2\times 10^{-3} \sigma^{-1}$ in the downstream rest frame in one-dimensional simulations \citep[][see also the Appendix B of this paper for consistent results from multidimensional simulations]{plotnikovsynchrotron2019M}.
This was soon verified in three-dimensional simulations, where an efficiency of $\sim 10^{-3} \sigma^{-1}$ in the downstream frame was found \citep{sironiCoherent2021PRL}.
}.

The configuration of magnetic fields near the shock front in GRB/FRB scenarios is another open question.
As the ejecta expands to a large distance, the frozen-in radial component of the magnetic field decays more rapidly with the radius ($\propto R^{-2}$) compared to the tangential component ($\propto R^{-1}$) due to magnetic flux conservation \citep{spruitLarge2001A}.
Therefore for simplicity, an ordered field perpendicular to the shock normal (perpendicular shock) is assumed in most theoretical works on shock dynamics \citep{kennelConfinement1984A, zhangGamma2005Aa, 2005JFM...544..323R, 2009ApJ...690L..47M, mimicaDeceleration2009A, 2010MNRAS.401..525M, narayanConstraints2011M, aimechanical2021M} and particle acceleration \citep{sironiMaximum2013A, iwamotoPersistence2017A, plotnikovsynchrotron2019M, sironiCoherent2021PRL}.

However, there could exist a tilted ordered field that makes an inclination angle with the shock normal (oblique shock) in realistic scenarios, e.g., an ejecta or a rotationally driven wind not too far away from the central engine.
In particular, the magnetic obliquity is important for diffusive shock acceleration in moderately magnetized shocks \citep[$10^{-3} \lesssim \sigma \lesssim 0.1$ ;][]{sironiRelativistic2015SSR}.
If the obliquity angle $\theta$ is larger than a critical angle $\theta_{\rm cr}$ (determined by the shock velocity in upstream rest frame), charged particles gyrating around the field lines would have to move faster than the speed of light along the field lines in order to outrun the shock front and return upstream \citep{kirkParticle1989M, begelmanShockDrift1990A, ballardFirstorder1991M}.
These shocks with $\theta > \theta_{\rm cr}$ are termed 'superluminal' while the opposite cases ($\theta < \theta_{\rm cr}$) are termed 'subluminal'.
It was found that the diffusive shock acceleration is significantly suppressed in moderately magnetized superluminal shocks \citep{sironiParticle2009A, sironiParticle2011A}.

Another field geometry that has been extensively discussed is a random field, which could be distorted by the instabilities, turbulence or magnetic reconnection in the flow \citep{medvedevGeneration1999A, zhangInternalcollisioninduced2011A, dengMagnetized2017A}.
An ordered field and a randomized field can be distinguished from their polarization signatures.
The ordered field generally yields a highly linearly polarized emission, whereas the random field leaves a low degree of polarization.
So far no conclusion regarding the polarization of GRB prompt emission has been reached from observations \citep[see review of][]{gillGRB2021a}.
The observed data of early optical afterglow emission suggested a high level of linear polarization, indicating that the reverse shock is dominated by an ordered field 
(GRB 090102, \citet{steeleTen2009N};
GRB 110205A, \citet{steelePolarimetry2017A};
GRB 120308A, \citet{mundellHighly2013N};
however see the low polarization in GRB 190114C, \citet{jordana-mitjansLowly2020A}).
The field configuration in forward shock emission was constrained for the radio afterglow of GRB 170817A associated with the binary neutron star coalescence GW170817, where a random field with a finite parallel component was favored \citep{gillAfterglow2018M, gillConstraining2020M}.

In contrast to the poor understanding of shock magnetization and field geometry from observations, the theoretical description of the shock has been well-established.
The jump conditions of relativistic oblique shocks were first presented in \citet{dehoffmannMagnetoHydrodynamic1950PRa}.
The mathematical structure of the shock equations was then detailed in \citet{1967rhm..book.....L} \citep[see also][]{1989rfmw.book.....A}.
As a special case, the solutions of the perpendicular shock jump conditions were systematically studied and applied to the pulsar wind problem \citep{kennelConfinement1984A} and the GRB problem \citep{zhangGamma2005Aa}.
Different methods were presented to solve the general jump conditions of oblique shocks \citep{webbRelativistic1987JPP, 1987PhFl...30.3045M, applShock1988A, ballardFirstorder1991M, 1999JPhG...25R.163K, summerlinDiffusive2012A}, including a modification of anisotropic pressure term motivated by particle acceleration simulations \citep{2004ApJ...600..485D}.
However, compared to the simple solutions of perpendicular shocks, the oblique shock jump conditions can only be solved numerically.
The methods to solve the jump conditions are often complex and lack of physical insights.
No attempt to find an asymptotic analytical solution has been made for oblique shocks.
Furthermore, previous studies only focus on  shocks with ordered magnetic fields, while the random field cases are insufficiently explored.

In this work, we revisit the jump conditions of relativistic oblique shocks with arbitrary magnetization.
Motivated by the astrophysical shocks in GRB/FRB scenarios, we explore shocks with an ordered field of an arbitrary obliquity and a random field of an arbitrary anisotropy parameter.
For oblique shocks that are at least mildly relativistic (relative Lorentz factor $\gamma_{\rm rel} \gtrsim 3$), we find that the jump conditions are determined by the tangential magnetization $\sigma_\perp$ (see the definition in equation~\eqref{eq:sigma_perp}) instead of the total magnetization $\sigma$.
In section~\ref{sec:general}, we describe the general jump conditions of the relativistic oblique shock.
An approximate solution is developed and compared to the strict solution in section~\ref{sec:quasinormal}.
In section~\ref{sec:randomshock}, we construct a shock model with random magnetic field and solve for the downstream parameters from jump conditions.
A summary of our findings is presented in section~\ref{sec:conclusions} with some discussions of future works.

\section{General solutions of relativistic magnetohydrodynamic shocks}
\label{sec:general}

The ideal relativistic magnetohydrodynamics is governed by conservation laws
\begin{align}
    & \nabla_\mu \left(\rho u^\mu\right) = 0\, ,\\
    & \nabla_\mu \left(b^\mu u^\nu - b^\nu u^\mu\right) = 0\, ,\\
    & \nabla_\mu \left(T^{\mu\nu}\right) = 0\, ,
\end{align}
where $T^{\mu\nu}$ is the energy-momentum tensor given by
\begin{equation}
    T^{\mu\nu} = \left(\rho h + b^2\right)u^\mu u^\nu + \left(p + \frac{b^2}{2}\right)g^{\mu \nu } - b^\mu b^\nu\, .
    \label{eq:tensor}
\end{equation}
Here, $\rho,\ p,\ h$ are the rest mass density, thermal pressure and specific enthalpy of the fluid, $u^\mu = (u^0, \boldsymbol{u}) = \gamma (1, \boldsymbol{\beta})$ is the four-velocity, $\gamma = 1/\sqrt{1-\beta^2}$ is the Lorentz factor, $\boldsymbol{\beta}$ is the three-vector of dimensionless velocity, $b^\mu = (b^0, \boldsymbol{b})$ is the magnetic field four-vector and $b^2 = b^\mu b_\mu$.
The specific enthalpy is related to $\rho$ and $p$ by the adiabatic index $\hat{\Gamma}$ through 
\begin{equation}
    \rho h = \rho c^2 + e + p = \rho c^2 + \frac{\hat{\Gamma}}{\hat{\Gamma}-1}p\, ,
\end{equation}
where $c$ is the speed of light and $e$ is the internal energy density.
In an arbitrary frame, the three-vectors of magnetic field $\boldsymbol{B}$ and electric field $\boldsymbol{E}$ are related to $b^\mu$ and $u^\mu$ by \citep{1999MNRAS.303..343K}
\begin{align}
    & \boldsymbol{B}/\sqrt{4\pi} = \boldsymbol{b} u^0 - \boldsymbol{u} b^0\, ,\\
    & \boldsymbol{E}/\sqrt{4\pi}c = \boldsymbol{b} \times \boldsymbol{u}\, ,
\end{align}
or inversely
\begin{align}
    & b^0 = \boldsymbol{u} \cdot \boldsymbol{B}/\sqrt{4\pi}\, ,\\
    & \boldsymbol{b} = \left(\boldsymbol{B}/\sqrt{4\pi} + b^0\boldsymbol{u}\right)/u^0\, .
\end{align}
Therefore in the fluid rest frame where $u^\mu = (1,0,0,0)$, we naturally get $b^\mu = (0,\boldsymbol{B}/\sqrt{4\pi})$ and $\boldsymbol{E} = 0$.

The discontinuity surface is a hypersurface defined by $\phi(x^\mu)=0$.
Let $l_\mu \equiv \partial_\mu \phi$ be the normal vector to the hypersurface, normalized such that $l^\mu l_\mu = 1$.
Then the jump conditions can be cast into
\begin{equation}
    \left[\left[\rho u^\mu\right]\right] l_\mu = \left[\left[b^\mu u^\nu - b^\nu u^\mu\right]\right] l_\mu = \left[\left[T^{\mu\nu}\right]\right] l_\mu = 0\, ,
\end{equation}
where $\left[\left[Q\right]\right]$ denotes the difference between upstream (region 1) value $Q_1$ and downstream  (region 2) value $Q_2$ of the same quantity $Q$, e.g., $\left[\left[Q\right]\right] \equiv Q_1 - Q_2$.
It is then convenient to define $u_l \equiv u^\mu l_\mu$, $b_l \equiv b^\mu l_\mu$, $V^\mu \equiv b^\mu u_l - b_l u^\mu$, $W^\mu \equiv (\rho h+b^2)u_l u^\mu + (p+b^2/2)l^\mu - b_l b^\mu$, and the jump conditions become
\begin{equation}
    \left[\left[\rho u_l\right]\right] = \left[\left[V^\mu\right]\right] = \left[\left[W^\mu\right]\right] = 0\, .
    \label{eq:general_jc}
\end{equation}

To solve the equations above, it is convenient to work in the rest frame of the shock (denoted as "s") in Cartesian coordinate, where we assume that the shock normal $l_\mu = (0,1,0,0)$ points along the x axis.
Throughout this paper, $Q_{ij}$ denotes the quantity $Q$ in region $i$ in the rest frame of $j$ ($i = 1,2$ and $j = 1,2,s$), following the symbol convention of \citet{zhangGamma2005Aa}.
Without loss of generality, we set the upstream four-velocity to be $u_{1s}^\mu = \gamma_{1s}(1,\beta_{1s}^x,0,0)$ such that $u_{l1} = u_{1s}^x = u_{1s}$ and assume the upstream magnetic field lines are confined to the $x-y$ plane.
Then it follows from the jump conditions~\eqref{eq:general_jc} that the velocity of the flow and the magnetic field downstream also lie within the $x-y$ plane \citep{applShock1988A}.
The downstream four-velocity is thus $u_{2s}^\mu = \gamma_{2s}(1,\beta_{2s}^x,\beta_{2s}^y,0)$ and $u_{l2} = u_{2s}^x$.

We assume a cold upstream flow where $e_1 = p_1 = 0$, which is the only assumption made here.
The degree of magnetization $\sigma$ and the obliquity angle $\theta$ (the angle between the magnetic field line and shock normal) are defined in the comoving frame, i.e.,
\begin{align}
    &\sigma_i \equiv \frac{B_i^2}{4\pi \rho_i h_i}\, ,\\
    &\theta_i \equiv \arcsin{\frac{B_{\perp i}}{B_i}}\, ,
\end{align}
where $B_i$ is the strength of magnetic field of region $i$ measured in the comoving frame.
The field components perpendicular and parallel to the shock normal are denoted as $B_\perp$ and $B_\parallel$, respectively.
The parameters in the upstream region are more fundamental, we therefore define
\begin{align}
    &\sigma \equiv \sigma_1 = \frac{B_1^2}{4\pi \rho_1 h_1}\, ,
    \label{eq:sigma}\\
    &\theta \equiv \theta_1 = \arcsin{\frac{B_{\perp 1}}{B_1}}\, .
\end{align}

It should be clarified here that in oblique shocks, this definition of the magnetization parameter is not, as what was presumed for perpendicular shocks, equivalent to the definition invoking flux ratio in the lab frame.
The other definition of magnetization is to take the ratio of the Poynting flux $T^{0x}_{\rm em}$ to the matter energy flux $T^{0x}_{\rm mat}$ given by the energy-momentum tensor~\eqref{eq:tensor}, i.e.,
\begin{equation}
    \sigma_{\rm flux} = \frac{T^{0x}_{\rm em}}{T^{0x}_{\rm mat}}
    = \frac{b^2 u^0 u^x - b^0 b^x}{\rho h u^0 u^x } = \frac{B^2 -(\boldsymbol{\beta} \cdot \boldsymbol{B}) B_x/\beta_x}{4\pi \rho h \gamma^2}\, ,
\end{equation}
where $T^{0x}_{\rm em}$ is the electromagnetic energy density flux in the x direction and $T^{0x}_{\rm mat}$ is the energy density flux of matter in the x direction.
Here, the magnetic field $\boldsymbol{B}$ and dimensionless velocity $\boldsymbol{\beta}$ are measured in the same frame.
In the shock rest frame, even if we assume the velocity is directed along the x direction such that $u^\mu = \gamma_{1s}(1,\beta_{1s},0,0),\, B_{\perp 1s} = \gamma_{1s}B_{\perp 1}$, we would get
\begin{equation}
    \sigma_{\rm flux} = \frac{B_{\perp 1s}^2}{4\pi \rho_1 h_1 \gamma_{1s}^2} = \frac{B_{\perp 1}^2}{4\pi \rho_1 h_1}\, ,
\end{equation}
which is not equivalent to the definition~\eqref{eq:sigma}.
Instead, this is exactly the tangential magnetization $\sigma_\perp$ we define in equation~\eqref{eq:sigma_perp}.
Since we find in this work that the jump conditions of relativistic shocks are governed by $\sigma_\perp$ instead of $\sigma$, the definition with flux ratio may be a better choice for magnetization parameter.
However, for clarification, we adopt the definition of magnetization~\eqref{eq:sigma} to distinguish it from $\sigma_\perp$.

We assume a cold upstream, and therefore the only adiabatic index that has physical meaning is the downstream one, which from now on we refer to as
\begin{equation}
    \hat{\Gamma} \equiv \hat{\Gamma}_2\, .
\end{equation}
Throughout this work, $\hat{\Gamma}=4/3$ is adopted.
A more general but simple approximation to the adiabatic index $\hat{\Gamma}$ is to consider the average microscopic motion of particles, which gives $\hat{\Gamma} \simeq (4\gamma_{21}+1)/3\gamma_{21}$ \citep[see equation  (4.20) of][]{zhangPhysics2018TPoGBbBZI9CUP}.
The adiabatic index thus goes to $5/3$ in the Newtonian limit and $4/3$ in the relativistic limit.
This expression has been adopted to discuss the oblique shock jump conditions \citep{2004ApJ...600..485D} as well as the mechanical model for blastwaves \citep{aimechanical2021M}.
Here, we are more interested in the relativistic regime and therefore take the relativistic limit $4/3$ as an approximation.

The jump conditions are thus fully determined by three dimensionless quantities in the upstream $(\sigma,\theta,u_{1s})$ if the downstream adiabatic index $\hat{\Gamma}$ is specified.
Using these four parameters as input, we can solve for the downstream dimensionless quantities numerically following the procedure presented in Appendix~\ref{sec:appendix_oblique} \citep[see also][for different treatments]{webbRelativistic1987JPP, 1987PhFl...30.3045M, applShock1988A, 1999JPhG...25R.163K}.

However, for an oblique shock with arbitrary obliquity angle, the equations governing the solutions to the jump conditions are often tedious and lack of physical insights.
It is difficult to arrive at an asymptotic analytical solution that hints at the essence of the problem.
Therefore, in the next section, we present a ``quasi-normal'' approximation to this problem, which proves to be a satisfying approximation when the shock is relativistic ($\gamma_{21} \gg 1$).

\section{Quasi-normal shocks: approximation for relativistic oblique shocks}
\label{sec:quasinormal}

Motivated by the early work of \citet{zhangGamma2005Aa} for perpendicular shock, we introduce the quasi-normal approximation, i.e., we assume the flow velocities are nearly directed along the shock normal ($\beta_{2s}^y \simeq 0$ in our formalism).

Consequently, we have $b^\mu_{is} \simeq (u_{is}B_{\parallel is},\gamma_{is}B_{\parallel is},B_{\perp is}/\gamma_{is})/\sqrt{4\pi}$ ($i=1,2$).
The six jump conditions thus lead to
\begin{equation}
\begin{split}
    &\left[\left[\rho u_l\right]\right] =
    \left[\left[B_{\parallel is}\right]\right] =
    \left[\left[\beta_{is}B_{\perp is}\right]\right] =
    \left[\left[\beta_{is}\left(\gamma_{is}^2 \rho h + \frac{B_{\perp is}^2}{4\pi}\right)\right]\right] \\
    &=
    \left[\left[u_{is}^2\rho h + p - \frac{B_{\parallel is}^2}{8\pi} + \left(1+\beta_{is}^2\right)\frac{B_{\perp is}^2}{8\pi}\right]\right] =
    \left[\left[B_{\parallel is}B_{\perp is}\right]\right]=
    0\, .
\end{split}
\end{equation}
It is clear that the second, third and sixth conditions can not hold simultaneously unless $B_{\parallel is}B_{\perp is}=0$.
This means a shock with flow velocities normal to the shock front would only be an exact solution for a strictly perpendicular shock ($B_\parallel = 0$) or parallel shock ($B_\perp = 0$).

However, as an approximation, we assume a relaxed sixth condition, where an extra term with a very small $u_{is}^y$ contributes to fulfill the sixth constraint.
We therefore assume the sixth condition is automatically satisfied and focus on the first five conditions.
They can be manipulated into four equations:
\begin{align}
     & \rho_1 u_{1s} = \rho_2 u_{2s}\, , \\
     & \mathcal{E} = \beta_{1s}B_{\perp 1s}=\beta_{2s}B_{\perp 2s}\, ,\\
     &\gamma_{1s}h_1+\frac{\mathcal{E} B_{\perp 1s}}{4\pi \rho_1 u_{1s}}=\gamma_{2s}h_2+\frac{\mathcal{E} B_{\perp 2s}}{4\pi \rho_2 u_{2s}}\, ,
     \label{eq:3}\\
     &u_{1s}h_1+\frac{p_1}{\rho_1 u_{1s}}+\frac{B_{\perp 1s}^2}{8\pi \rho_1 u_{1s}}= u_{2s}h_2+\frac{p_2}{\rho_2 u_{2s}}+\frac{B_{\perp 2s}^2}{8\pi \rho_2 u_{2s}}\, ,
     \label{eq:4}
\end{align}
which are identical to equations (1)-(4) of \citet{zhangGamma2005Aa} but supplemented by $B_{\parallel 1s}=B_{\parallel 2s}$, i.e. the parallel component of magnetic field is decoupled from the rest of the relations.

It is thus convenient to define a tangential degree of magnetization
\begin{equation}
    \sigma_\perp \equiv \sigma \sin^2 \theta = \frac{B_{\perp 1}^2}{4\pi \rho_1 h_1}\, .
    \label{eq:sigma_perp}
\end{equation}
Once $(\sigma_\perp,\gamma_{21},\hat{\Gamma})$ is specified, the jump conditions can be solved following the method of \citet{zhangGamma2005Aa}, but with $\sigma$ in their equations replaced by $\sigma_\perp$.
However, \citet{zhangGamma2005Aa} solves for the downstream four-velocity $u_{2s}=u_{2s}(\sigma,\gamma_{21},\hat{\Gamma})$.
Here, we find it more convenient to solve for the downstream dimensionless velocity $\beta_{2s}=\beta_{2s}(\sigma_\perp,\gamma_{21},\hat{\Gamma})$ following the method described below.

Within the quasi-normal approximation, the Lorentz transformation of magnetic field yields $B_{\parallel is}=B_{\parallel i}, B_{\perp is} = \gamma_{is}B_{\perp i}$.
The obliquity angles in region $1$ and region $2$ are related to each other by exploiting the continuity of parallel magnetic field $B_{\parallel 1}=B_{\parallel 2}$ at the shock front, which gives a proper compression ratio
\begin{equation}
    R \equiv \frac{\rho_2}{\rho_1}=\frac{u_{1s}}{u_{2s}}=\frac{B_{\perp 2}}{B_{\perp 1}}=\frac{\tan \theta_2}{\tan \theta_1}\, .
\end{equation}
The compression ratio is thus related to the proper compression ratio by
\begin{equation}
    r \equiv \frac{\beta_{1s}}{\beta_{2s}}=\frac{\gamma_{2s}}{\gamma_{1s}}R=\frac{B_{\perp 2s}}{B_{\perp 1s}}\, .
\end{equation}
With some algebra (see Appendix~\ref{sec:appendix_quasinormal}), by defining a hydrodynamical solution
\begin{equation}
    \beta_{2s,0} \equiv (\hat{\Gamma}-1)\frac{\beta_{21}}{1+\sqrt{1-\beta_{21}^2}}\, ,
\end{equation}
we can solve for $\beta_{2s}=\beta_{2s}(\sigma_\perp,\gamma_{21},\hat{\Gamma})$ by solving a three-order equation
\begin{equation}
    (\beta_{2s}-\beta_{2s,0})\beta_{2s}(\beta_{2s}+\beta_{21})=\left[\beta_{2s}+\left(1-\frac{\hat{\Gamma}}{2}\right)\beta_{21}\right](1-\beta_{2s}^2)\sigma_\perp\, .
    \label{eq:threeorder}
\end{equation}

The other quantities are thus fully determined by $\sigma_\perp,\gamma_{21},\hat{\Gamma}$ when $\beta_{2s}$ is known.
We list some of the quantities below:
\begin{align}
    &u_{2s} = \gamma_{2s}\beta_{2s}\, ,
    \label{eq:quant1}\\
    &\frac{\gamma_{1s}}{\gamma_{21}} = \gamma_{2s}(1+\beta_{2s}\beta_{21})\, ,
    \label{eq:quant2}\\
    &\frac{e_2}{\rho_2 c^2} = (\gamma_{21}-1)\left[1-\frac{1}{2}\frac{\gamma_{21}\beta_{21}^2}{\gamma_{21}-1}\frac{\beta_{2s}-\beta_{2s,0}}{\beta_{2s}+\left(1-\frac{\hat{\Gamma}}{2}\right)\beta_{21}}\right]\, ,
    \label{eq:quant3}\\
    &\frac{\rho_2}{\rho_1} = \frac{B_{\perp 2}}{B_{\perp 1}} = R = \gamma_{21}\left(1+\frac{\beta_{21}}{\beta_{2s}}\right)\, ,
    \label{eq:quant4}\\
    &\frac{p_{b\perp 2}}{p_2} = \frac{R\sigma_\perp}{2(\hat{\Gamma}-1)}\left(\frac{e_2}{\rho_2 c^2}\right)^{-1}
    \label{eq:quant5}
    \, ,
\end{align}
where $p_{b\perp} \equiv B_\perp^2/8\pi$ is the magnetic pressure attributed to the tangential field component.

We discuss two asymptotic regimes in section~\ref{sec:subhydro} and section~\ref{sec:subrelativistic}.
In section~\ref{sec:subgeneral}, we compare the quasi-normal approximation with the exact solutions solved numerically in section~\ref{sec:general}.

\subsection{Hydrodynamic or parallel shock ($\sigma_\perp = 0$)}
\label{sec:subhydro}

It has been proved in the early work of \citet{dehoffmannMagnetoHydrodynamic1950PRa} that the jump conditions of a strictly parallel shock are the same as hydrodynamic shock.
With $\sigma_\perp=0$, the solution to equation~\eqref{eq:threeorder} is simply the hydrodynamic solution
\begin{equation}
    \beta_{2s} = (\hat{\Gamma}-1)\frac{\beta_{21}}{1+\sqrt{1-\beta_{21}^2}}\, .
\end{equation}
Substitute this solution into equations~\eqref{eq:quant1}\eqref{eq:quant2}\eqref{eq:quant3}\eqref{eq:quant4}, and we retrieve the hydrodynamic solutions presented in \citet{blandfordFluid1976PF}.
The solutions were summarized in equations (12)-(16) of \citet{zhangGamma2005Aa}.

\subsection{Ultra-relativistic shock ($\gamma_{21} \gg 1$)}
\label{sec:subrelativistic}

In the highly relativistic limit where the relative Lorentz factor $\gamma_{21} \gg 1$, we have $\beta_{21} \to 1$.
The governing equation~\eqref{eq:threeorder} reduces to a quadratic equation
\begin{equation}
    \beta_{2s}^2-\left[\hat{\Gamma}-1+(1-\hat{\Gamma}/2)\frac{\sigma_\perp}{1+\sigma_\perp} \right]\beta_{2s}-(1-\hat{\Gamma}/2)\frac{\sigma_\perp}{1+\sigma_\perp} = 0\, .
\end{equation}
The solution is simply
\begin{equation}
    \beta_{2s} = \left(\hat{\Gamma}-1+X+\sqrt{(\hat{\Gamma}-1+X)^2+4X}\right)/2\, ,
    \label{eq:asymptotic}
\end{equation}
where $X = (1-\hat{\Gamma}/2)\sigma_\perp/(1+\sigma_\perp)$.
This is the ultra-relativistic solution found by \citet{kennelConfinement1984A} and retrieved in \citet{zhangGamma2005Aa} for strictly perpendicular shock.
Here, we show that this asymptotic solution also holds in the general picture of oblique shocks with arbitrary obliquity, but with $\sigma$ replaced by $\sigma_\perp$.
In the next subsection, we prove that this ultra-relativistic limit is a good approximation to the exact solution for $\gamma_{21} \gtrsim 10$.

For an ultra-relativistic shock, the adiabatic index is close to the relativistic limit $\hat{\Gamma} \simeq 4/3$.
It is therefore clear that $1/3 \le \beta_{2s} < 1$, where the lower limit is obtained for hydrodynamic case and the upper limit for strongly magnetized case.
Specifically for an ultra-relativistic strongly magnetized shock ($\gamma_{21} \gg 1,\, \sigma_\perp \gg 1$), we get
\begin{align}
    & \gamma_{2s} \simeq u_{2s} \simeq \sqrt{\sigma_\perp}\, ,\\
    & \gamma_{1s} \simeq 2\gamma_{21}\sqrt{\sigma_\perp}\, ,\\
    & e_2/\rho_2 c^2 \simeq 3\gamma_{21}/4\, ,\\
    & \rho_2/\rho_1 \simeq B_{\perp 2}/B_{\perp 1} \simeq 2\gamma_{21}\, .
\end{align}

\subsection{The general cases}
\label{sec:subgeneral}

\begin{figure*}
\centering
\begin{subfigure}{0.43\textwidth}
    \includegraphics[width=\textwidth]{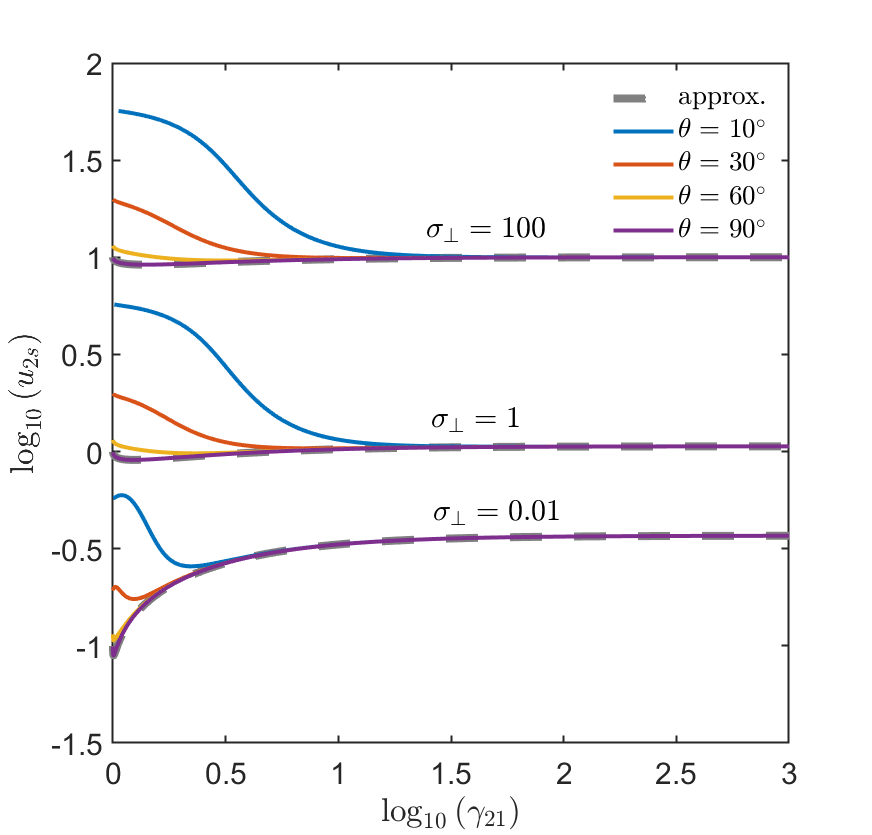}
\end{subfigure}
\begin{subfigure}{0.43\textwidth}
    \includegraphics[width=\textwidth]{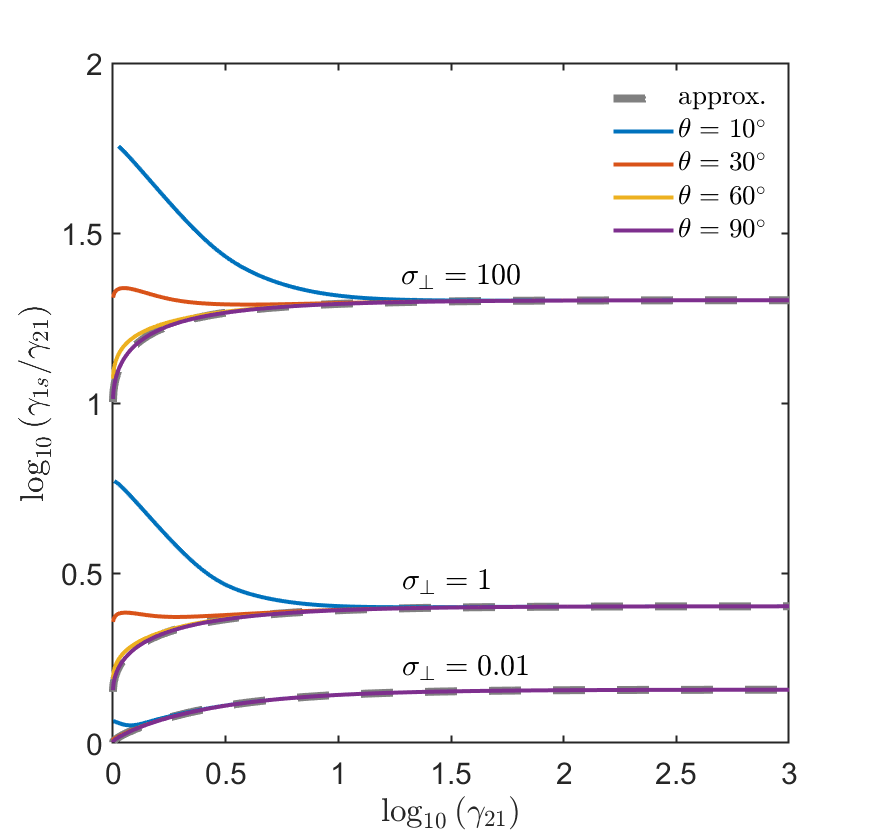}
\end{subfigure}
\begin{subfigure}{0.43\textwidth}
    \includegraphics[width=\textwidth]{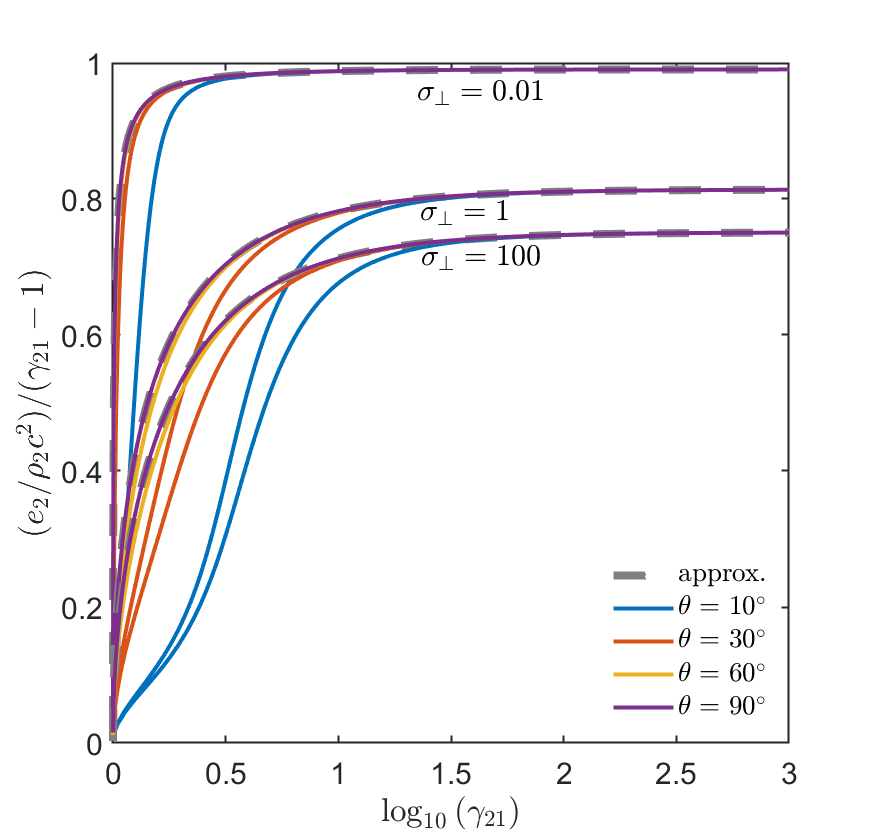}
\end{subfigure}
\begin{subfigure}{0.43\textwidth}
    \includegraphics[width=\textwidth]{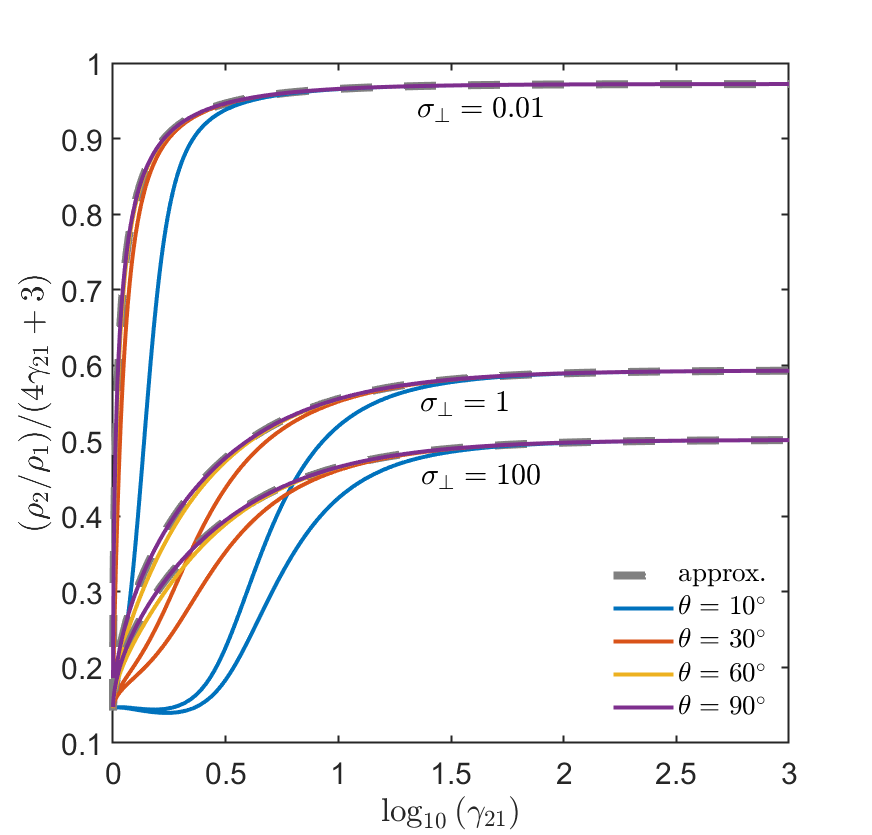}
\end{subfigure}
\begin{subfigure}{0.43\textwidth}
    \includegraphics[width=\textwidth]{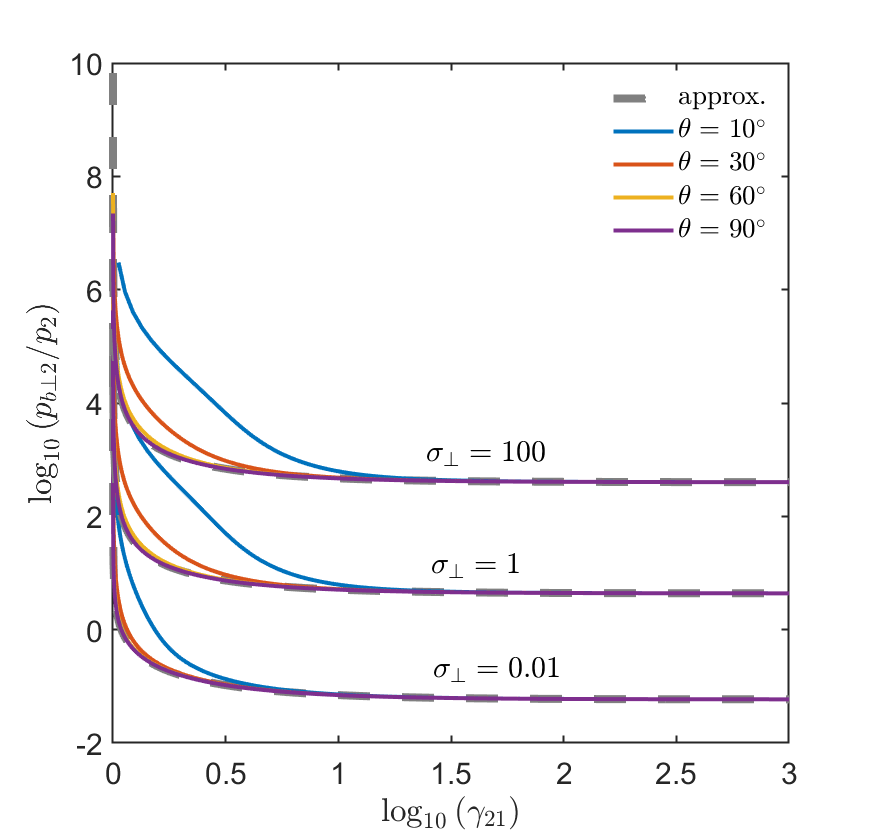}
\end{subfigure}
\begin{subfigure}{0.43\textwidth}
    \includegraphics[width=\textwidth]{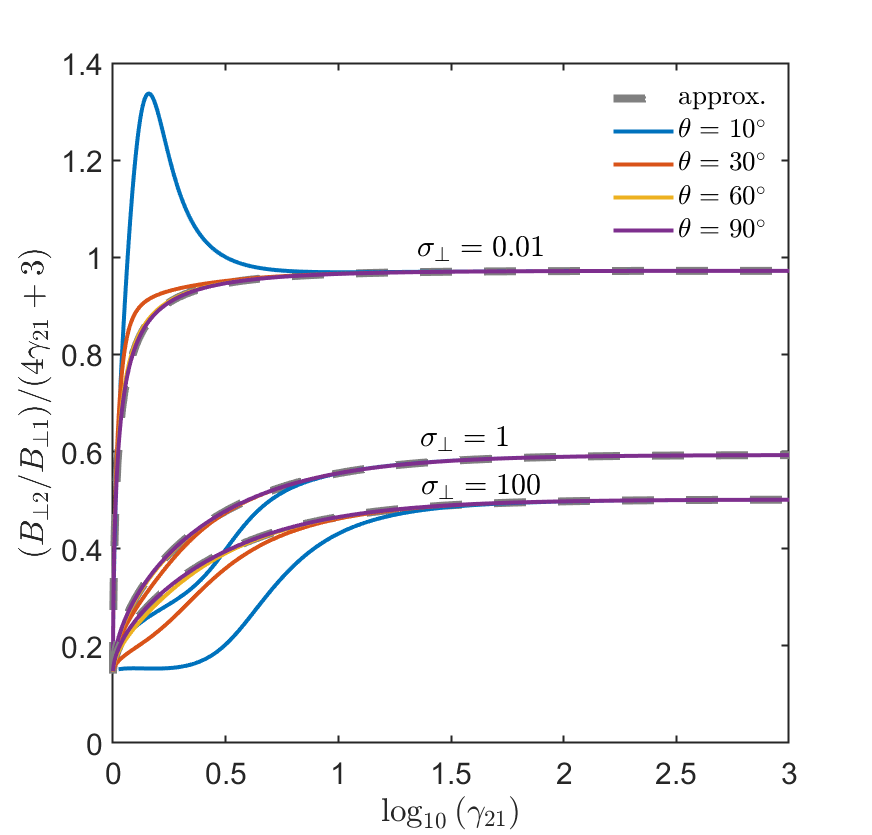}
\end{subfigure}
    \caption{
    Shock parameters $u_{2s},\, \gamma_{1s}/\gamma_{21},\, e_2/\rho_2 c^2,\, \rho_2/\rho_1,\, p_{b\perp 2}/p_2,\, B_{\perp 2}/B_{\perp 1}$ as functions of relative Lorentz factor $\gamma_{21}$, for given tangential magnetization $\sigma_\perp$ and obliquity angle $\theta$ of an ordered magnetic field.
    The grey dashed lines represent solutions given by quasi-normal approximation.
    The colored solid lines indicate exact solutions to oblique shock jump conditions, with different obliquity angles $\theta$ denoted by different colors.
    The exact solutions with different obliquity but the same $\sigma_\perp$ converge to the same approximated solution, as labeled by the value of $\sigma_\perp$.
    The parameters $e_2/\rho_2 c^2$, $\rho_2/\rho_1$ and $B_{\perp 2}/B_{\perp 1}$ are normalized with respect to their hydrodynamic values, in accordance with \citet{zhangGamma2005Aa}.
	}
    \label{fig:1}
\end{figure*}

For arbitrary values of $\sigma_\perp$ and $\gamma_{21}$, the three-order equation~\eqref{eq:threeorder} can only be solved numerically.
The solutions are the same with the results presented in \citet{zhangGamma2005Aa}, but with $\sigma$ in their equations replaced by $\sigma_\perp$.

In Fig.~\ref{fig:1}, we present the quasi-normal approximation given by equation~\eqref{eq:threeorder} (grey dashed lines) and the exact solutions to oblique shock jump conditions in section~\ref{sec:general} (colored solid lines).
The exact solutions with different obliquity angles $\theta = 10^\circ,\, 30^\circ,\, 60^\circ,\, 90^\circ$ (shown with different colors as denoted in the legend) but the same tangential magnetization $\sigma_\perp$ converge to the same approximated solution (marked by the value of $\sigma_\perp$) as $\gamma_{21}$ increases.
Six parameters $u_{2s},\, \gamma_{1s}/\gamma_{21},\, e_2/\rho_2 c^2,\, \rho_2/\rho_1,\, p_{b\perp 2}/p_2,\, B_{\perp 2}/B_{\perp 1}$ are selected to show the variations of solutions with $\gamma_{21}$, $\sigma_\perp$ and $\theta$, in accordance with \citet{zhangGamma2005Aa}
\footnote{
For the last parameter, we choose the amplification of tangential field component $B_{\perp 2}/B_{\perp 1}$ instead of the $F$ parameter in \citet{zhangGamma2005Aa} for the following reasons.
The $F$ parameter was created in the context of GRB afterglow model to determine whether the reverse shock is relativistic, which assumes perpendicular shocks.
However, for oblique shocks there could be more than two shocks when two flows of plasma collide into each other \citep{1999MNRAS.303..343K, 2006JFM...562..223G}, which makes it difficult to determine whether the reverse shock is relativistic prior to solving the problem (for a discussion of this point, see section~\ref{sec:conclusions}).
Furthermore, the assumption that the forward shock region is strictly hydrodynamical in \citet{zhangGamma2005Aa} would contradict the jump conditions at contact discontinuity if the shock is not strictly perpendicular (for a discussion of this point, see Appendix~\ref{sec:appendix_cd}).
Therefore, the parameter $F$ loses its physical meaning for oblique shocks.
}.
The parameters $e_2/\rho_2 c^2$, $\rho_2/\rho_1$ and $B_{\perp 2}/B_{\perp 1}$ are normalized to their values in hydrodynamic case.

The exact solutions for a $90^\circ$ shock (purple solid lines) coincide with our approximation (grey dashed lines) in all the panels, which verifies that the quasi-normal approximation becomes the exact solution for perpendicular shocks.
For all the parameters shown here, the deviations of the exact solutions from the quasi-normal approximation increase as the obliquity angle decreases when $\gamma_{21} \lesssim 10$.
In the relativistic regime ($\gamma_{21} \gtrsim 10$), however, all of the exact solutions are insensitive to the obliquity angle and converge to the approximated values given by the asymptotic analytical solutions~\eqref{eq:asymptotic}.
This can be easily understood in terms of the Lorentz boost of the tangential field component \citep[e.g.,][]{gallantRelativistic1992A, lemoineCorrugation2016A, plotnikovPerpendicular2018M}.
In the upstream where we assume the flow velocity is normal to the shock front, the tangential field component $B_{\perp 1s} = \gamma_{1s} B_{\perp 1}$ in the shock frame is boosted via the Lorentz transformation, whilst the parallel component $B_{\parallel 1s} = B_{\parallel 1}$ remains essentially the same.
The Lorentz factor $\gamma_{1s}$ is at least of the same order of magnitude as $\gamma_{21}$.
Consequently, for an ultra-relativistic shock ($\gamma_{21} \gg 1$), the magnetic field in the shock frame is dominated by the tangential component unless the shock is subluminal and near parallel ($\sin \theta \lesssim 1/\gamma_{1s}$).
Since the jump conditions are written in the shock frame, we expect that the tangential magnetization governs the jump conditions for ultra-relativistic shocks with moderate obliquity\footnote{However, this qualitative argument is not adequate to determine how well the quasi-normal approximation fits with the exact solutions when the shock transits from ultra-relativistic to mildly relativistic regime.}.

Here, we find that as long as the shock is at least mildly relativistic ($\gamma_{21} \gtrsim 3$), the quasi-normal approximation proves to be a good approximation, either for weakly magnetized shocks ($\sigma \lesssim 10^{-2}$) with arbitrary obliquity or for moderately oblique shocks ($\theta \gtrsim 30^\circ$) with arbitrary magnetization.
In that sense, we find the jump conditions of relativistic oblique shocks are governed by tangential magnetization $\sigma_\perp$ instead of the total magnetization $\sigma$.
As found by \citet{zhangGamma2005Aa}, both $e_2/\rho_2 c^2$ and $\rho_2/\rho_1$ are suppressed by strong magnetic field, but approach the ultra-relativistic strongly magnetized limit $e_2/\rho_2 c^2 \simeq 3\gamma_{21}/4,\, \rho_2/\rho_1 \simeq 2\gamma_{21}$ given in section~\ref{sec:subrelativistic}.
The amplification of tangential field component $B_{\perp 2}/B_{\perp 1}$ exceeds the density ratio for near-Newtonian shocks with small obliquity angles, in agreement with \citet{applShock1988A}.
Nevertheless, the order of magnitude of the amplification factor can be well-estimated by $B_{\perp 2}/B_{\perp 1} \simeq 4\gamma_{21}$ for all cases as long as the shock is at least mildly relativistic, which is only suppressed by a factor of two for strongly magnetized shocks.

\begin{figure}
\centering
\begin{subfigure}{0.9\columnwidth}
    \includegraphics[width=\textwidth]{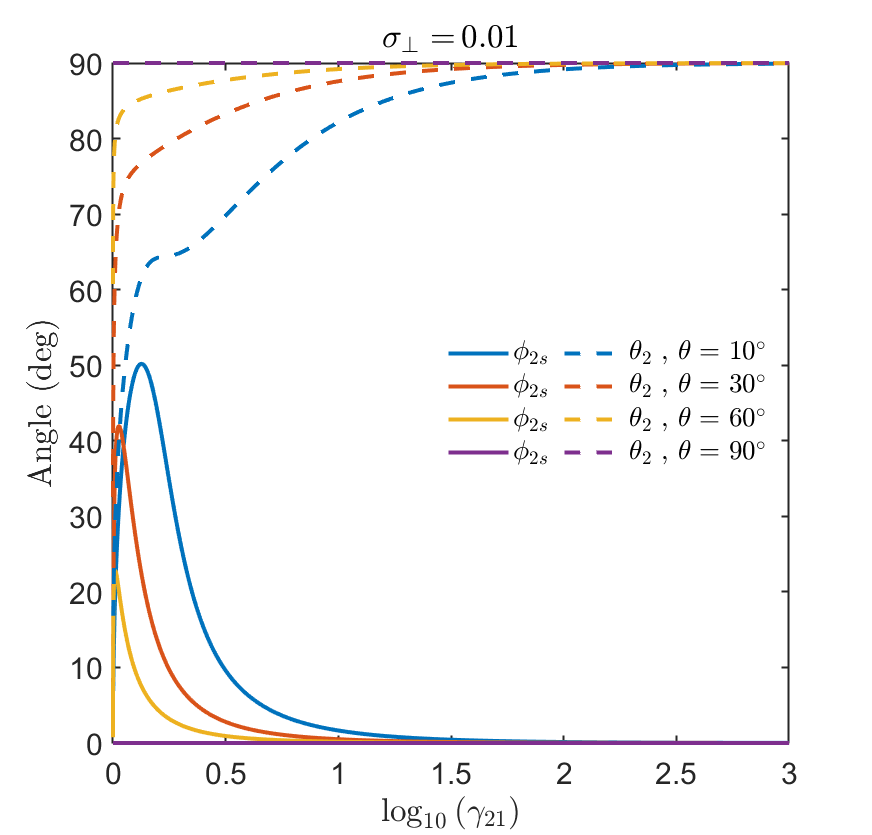}
\end{subfigure}
\begin{subfigure}{0.9\columnwidth}
    \includegraphics[width=\textwidth]{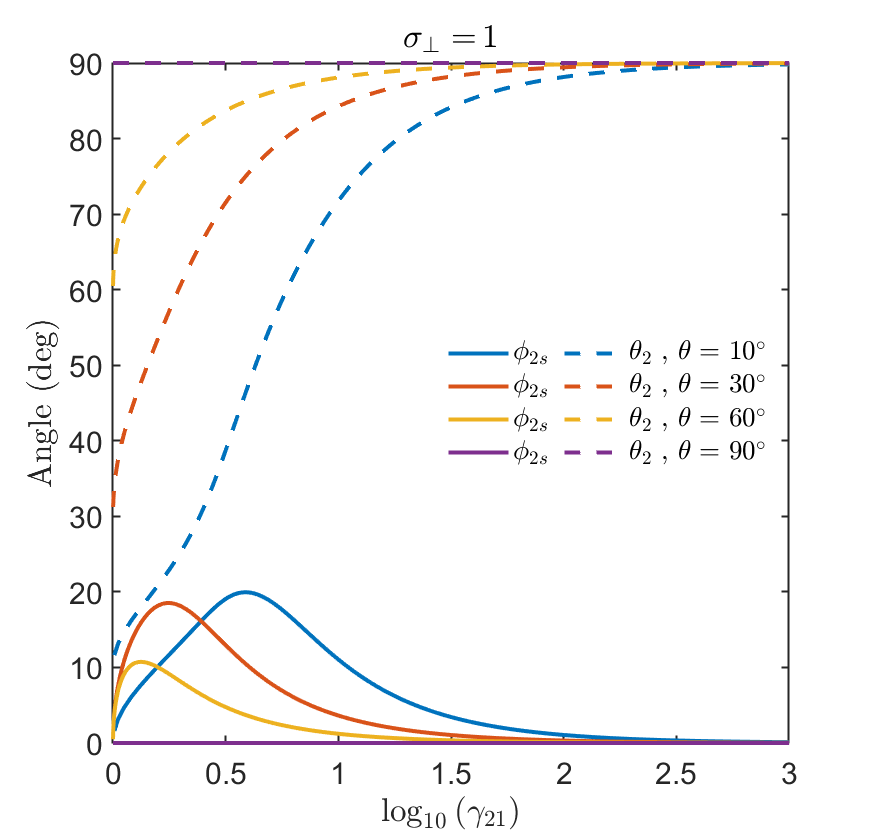}
\end{subfigure}
\begin{subfigure}{0.9\columnwidth}
    \includegraphics[width=\textwidth]{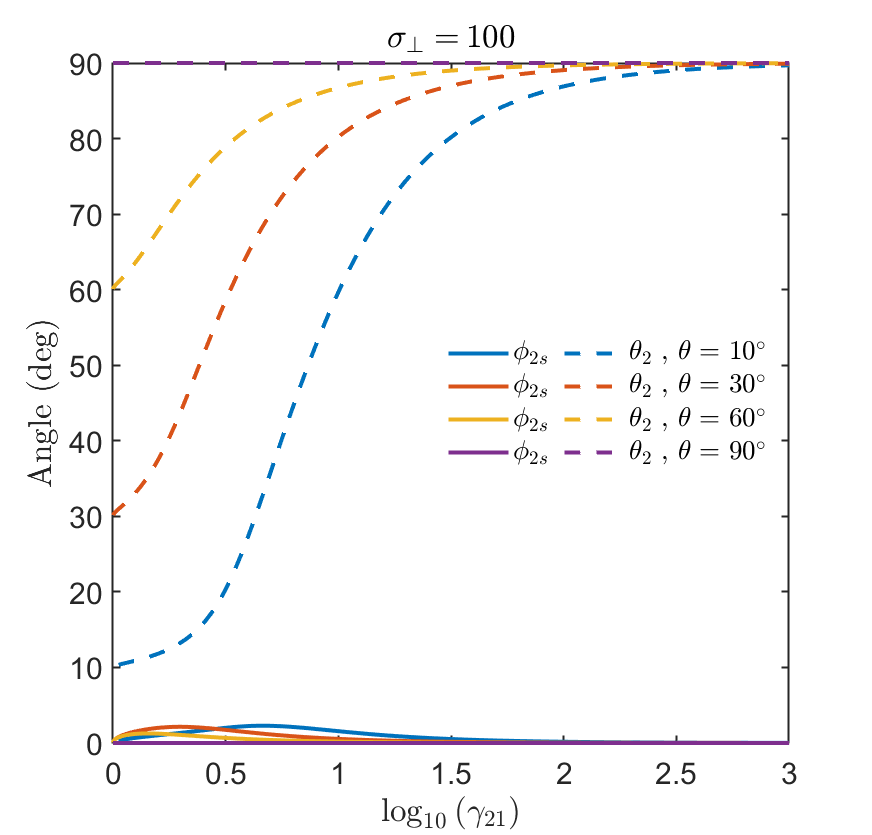}
\end{subfigure}
	\caption{
	Downstream inclination angles of velocity $\phi_{2s}$ and inclination angles of magnetic field $\theta_2$, as functions of relative Lorentz factor $\gamma_{21}$, for given tangential magnetization $\sigma_\perp$ and obliquity angle $\theta$ of an ordered magnetic field.
	The upper, middle and lower panels show the variations of angles for $\sigma_\perp=0.01,\, 1,\, 100$, respectively.
	The solid lines are for $\phi_{2s}$ while the dashed lines are for $\theta_2$, with different obliquity angles $\theta$ denoted by different colors.
	}
	\label{fig:2}
\end{figure}

The exact solutions for relativistic oblique shocks can be well-approximated by quasi-normal shocks because the downstream flow is nearly directed along the shock normal when the shock is relativistic.
This point can be directly seen from Fig.~\ref{fig:2}, where we plot the variations of downstream inclination angle of velocity $\phi_{2s} = \arctan{\beta_{2s}^y/\beta_{2s}^x}$ measured in the shock frame (solid lines) and downstream inclination angle of magnetic field $\theta_2$ measured in the fluid rest frame (dashed lines).
For fixed $\sigma_\perp$ and $\theta$, as the relative Lorentz factor $\gamma_{21}$ increases, the velocity inclination angle $\phi_{2s}$ reaches a peak in trans-relativistic regime, and then rapidly declines to nearly zero.
This trend is consistent with the result of \citet{applShock1988A}.
The maximum value of $\phi_{2s}$ can be higher than $45^\circ$ for near-Newtonian weakly magnetized shocks with high obliquity, but this value is greatly suppressed with increasing $\sigma_\perp$.
Fig.~\ref{fig:2} also shows that the post-shock obliquity angle $\theta_2$ is significantly enhanced to nearly $90^\circ$ by the relativistic shock due to the scaling of $B_{\perp 2}/B_{\perp 1} \simeq 4\gamma_{21}$.
For weakly magnetized shocks, it is thus reasonable to assume the shock-compressed magnetic field is nearly confined to the shock plane.
In strongly magnetized shocks, the amplification of tangential field component is weakly suppressed by a factor of two, and thus we find the post-shock obliquity angle $\theta_2$ climbs more slowly with increasing $\gamma_{21}$ for higher $\sigma_\perp$.
For a moderately magnetized quasi-parallel shock ($\sigma_\perp \sim 1,\, \theta \lesssim 10^\circ$), the shock-compressed field may not be dominated by the tangential component if the shock is only mildly relativistic.

\section{Relativistic shocks with random magnetic fields}
\label{sec:randomshock}

\subsection{A model for anisotropic random magnetic fields}

For a sufficiently small region near the shock, the magnetic field can be treated as a locally ordered field with uniform distribution.
However, for a field that is not globally ordered, its strength and inclination angle could have a statistical distribution at the shock front.
We therefore assume a field of different strengths and orientations at different regions near the shock front, but is locally uniform at each small region.
The magnetic field is thus locally divergence-free but globally randomized to some extent, and for simplicity we assume the distribution is axisymmetric with respect to the shock normal.

A general description of an anisotropic distribution of the random magnetic field is to allow the field strength to vary with inclination angle through $B(\theta)$ and a probability function $f(\theta)$ for the field to be in the specific solid angle \citep{sariLinear1999A}.
The probability to find a field of strength $B=B(\theta)$ within the interval $[\theta,\theta+d\theta]$ is thus $f(\theta)\sin\theta d\theta$.
One way to construct an anisotropic random field, as proposed by \citet{sariLinear1999A}, is to first take an isotropic distribution (with $B(\theta)$ and $f(\theta)$ both being constant) and then stretch it along the shock normal
\footnote{This is analogous to transforming a sphere into an ellipsoid with the semi-major axis $\xi$ times larger than the semi-minor axis.}, such that the maximum parallel field component $B_{\parallel\max}$ equals the maximum perpendicular field component $B_{\perp\max}$ multiplied by an parameter $\xi$ for anisotropy, i.e.,
\begin{equation}
    \xi \equiv \frac{B_{\parallel\max}}{B_{\perp\max}}\, .
\end{equation}
This yields \citep{sariLinear1999A, gillConstraining2020M}
\begin{align}
    & B(\theta)=\frac{B_{\perp\max}}{\sqrt{\sin^2\theta+\cos^2\theta/\xi^2}}\, ,
    \label{eq:bfieldfunc}\\
    & f(\theta)=\frac{1}{2\xi\left(\sin^2\theta+\cos^2\theta/\xi^2\right)^{3/2}}\, .
\end{align}
From now on, we use the notation $\left\langle\right\rangle$ to denote the value weighted averaged over all solid angles, such that $\left\langle Q \right\rangle \equiv \int_0^\pi Q f(\theta)\sin\theta d\theta$.
Some useful relations can be derived through simple algebra, e.g., \citep{gillConstraining2020M}
\begin{equation}
    \left\langle B_\perp^2 \right\rangle = \frac{2}{3} B_{\perp\max}^2\, ,\
    \left\langle B_\parallel^2 \right\rangle = \frac{\xi^2}{3} B_{\perp\max}^2\, ,\
    \left\langle B^2 \right\rangle = \frac{2+\xi^2}{3} B_{\perp\max}^2\, .
\end{equation}
Or in terms of another parameter
\footnote{From now on, we use $b$ to denote the parameter of field anisotropy defined by equation~\eqref{eq:littleb}.
This is not to be confused with the magnitude of the four-vector of magnetic field in Section~\ref{sec:general}.}
\begin{equation}
    b \equiv \frac{2\left\langle B_\parallel^2 \right\rangle}{\left\langle B_\perp^2 \right\rangle}
    \label{eq:littleb}
\end{equation}
to parameterize the field anisotropy for the discussion of polarization \citep{sariLinear1999A,gillConstraining2020M,gillLinear2020M,teboulImpact2021M}, we find
\begin{equation}
    b = \xi^2\, .
\end{equation}
The special case of $b=\xi=1$ corresponds to a completely random field with isotropic distribution.
When $b,\xi \ll 1$, the model describes a fully random field confined to the shock plane, e.g., generated via the relativistic two-stream instability \citep{medvedevGeneration1999A}.
The opposite extreme $b,\xi \gg 1$ represents an ordered field along the shock normal. For a visualized sketch of the field distribution, see Figure 2 in \citet{gillConstraining2020M}.

\subsection{Jump conditions of relativistic shocks with random magnetic fields}

\begin{figure*}
\centering
\begin{subfigure}{0.43\textwidth}
    \includegraphics[width=\textwidth]{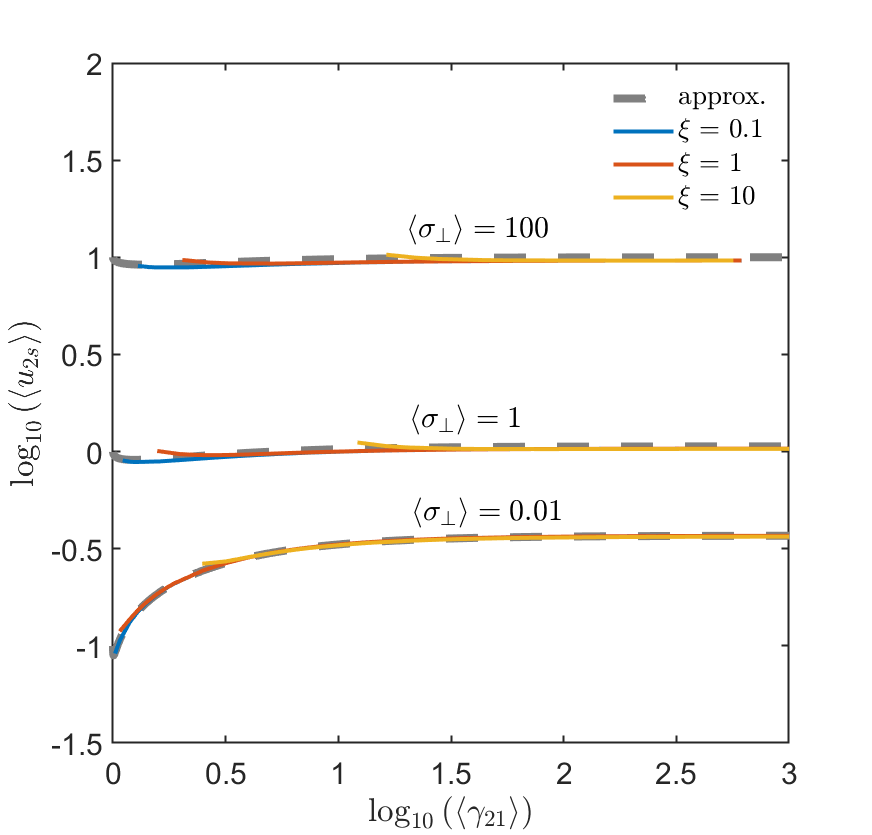}
\end{subfigure}
\begin{subfigure}{0.43\textwidth}
    \includegraphics[width=\textwidth]{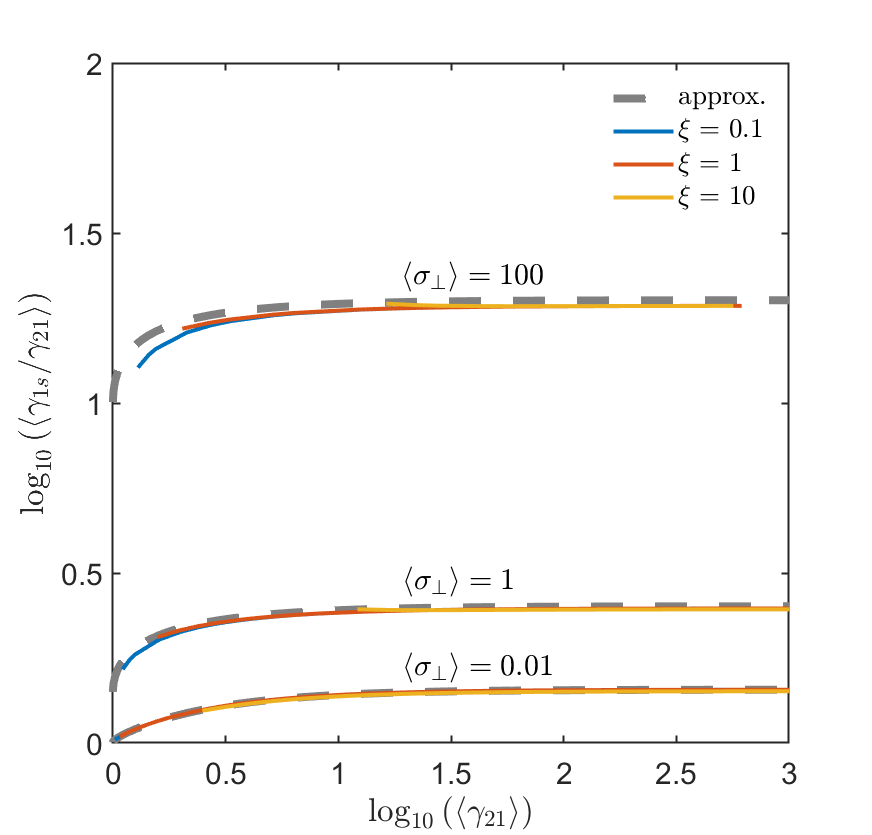}
\end{subfigure}
\begin{subfigure}{0.43\textwidth}
    \includegraphics[width=\textwidth]{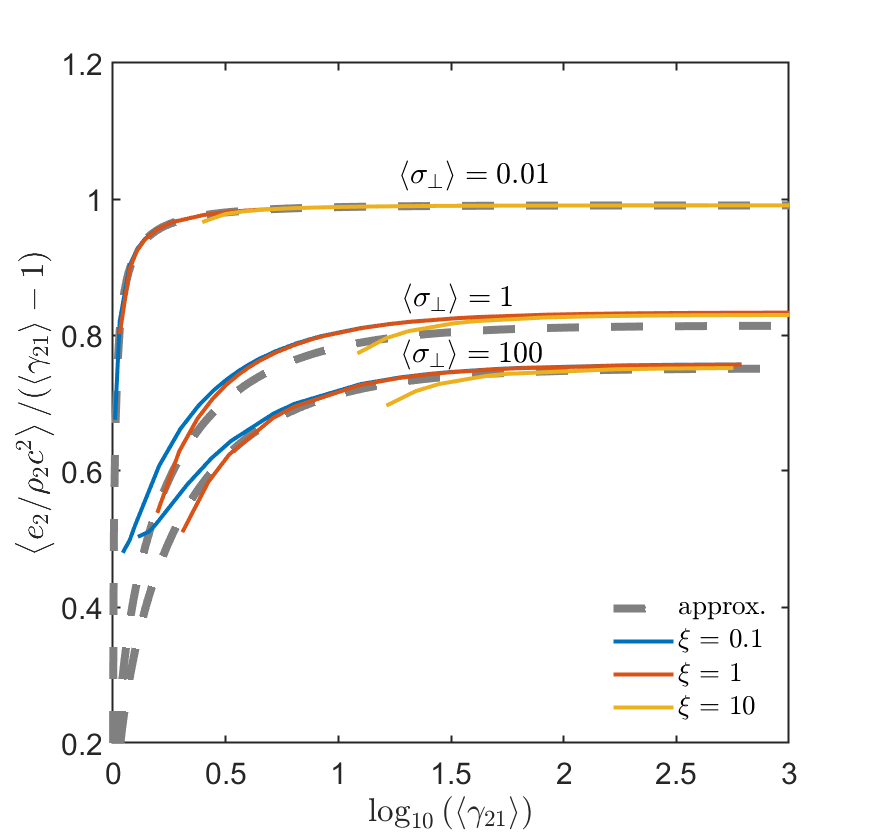}
\end{subfigure}
\begin{subfigure}{0.43\textwidth}
    \includegraphics[width=\textwidth]{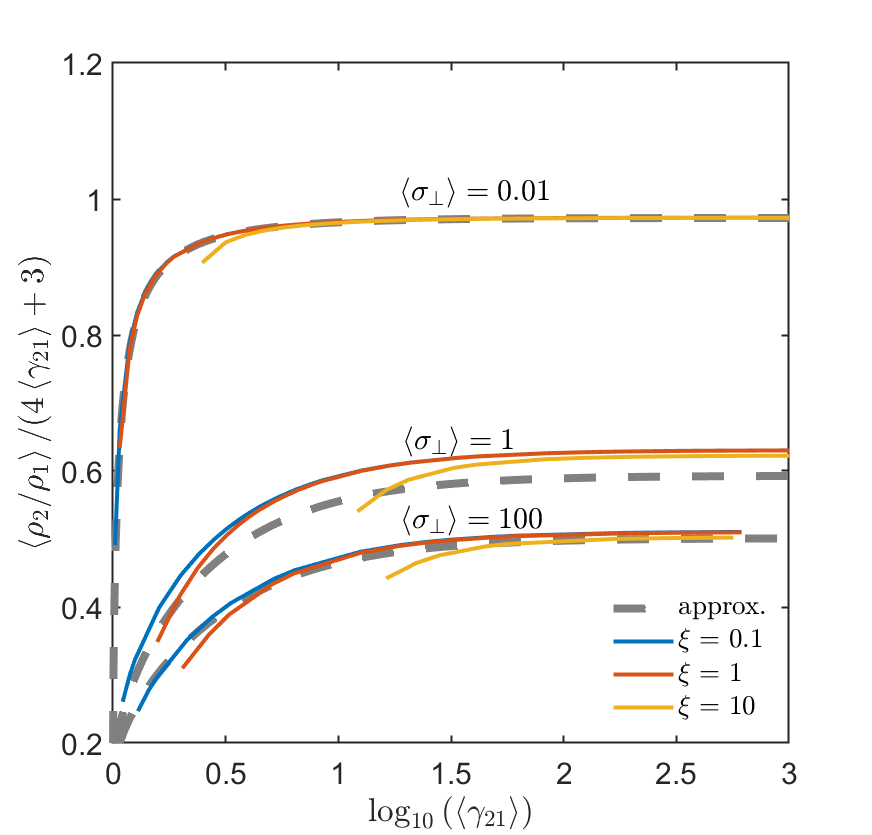}
\end{subfigure}
\begin{subfigure}{0.43\textwidth}
    \includegraphics[width=\textwidth]{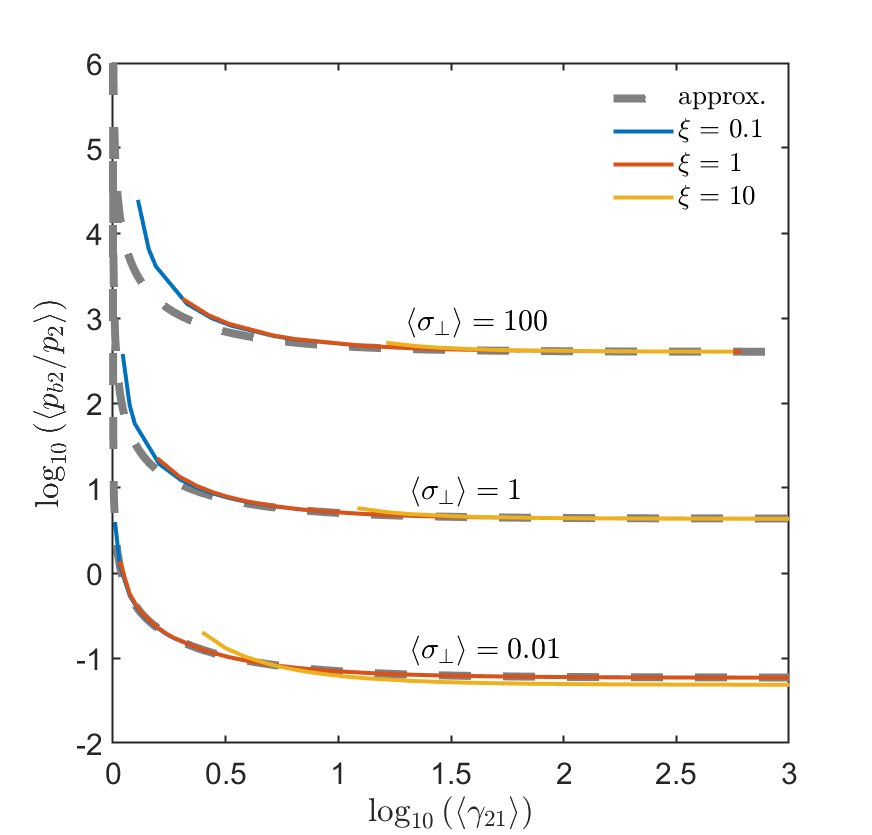}
\end{subfigure}
\begin{subfigure}{0.43\textwidth}
    \includegraphics[width=\textwidth]{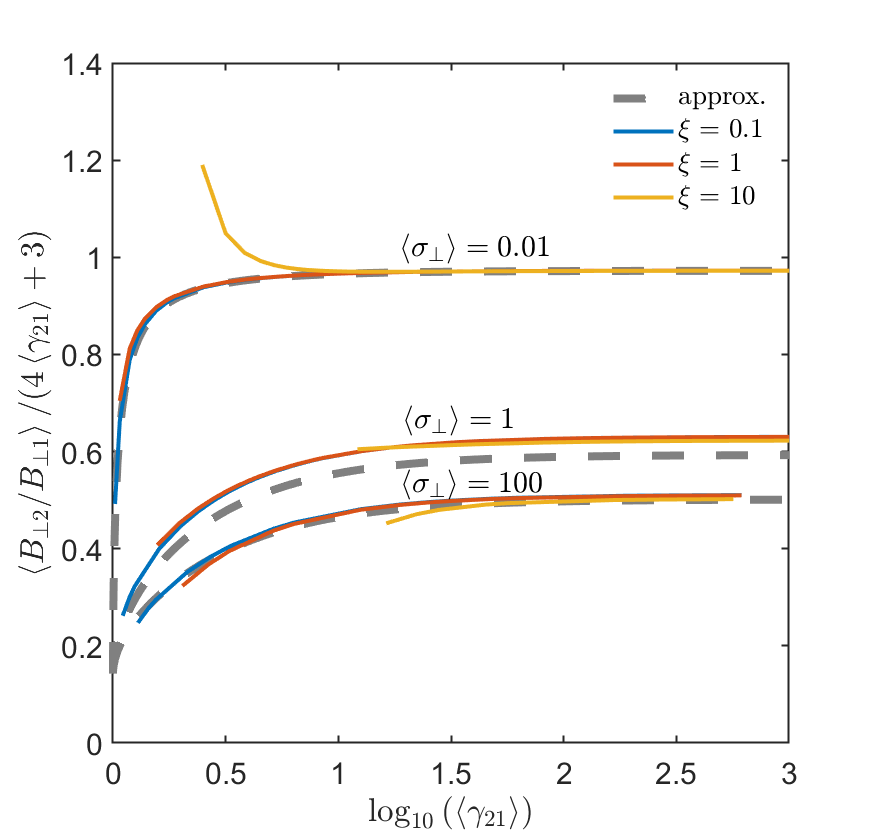}
\end{subfigure}
    \caption{
    Average shock parameters $\left\langle u_{2s}\right\rangle,\, \left\langle \gamma_{1s}/\gamma_{21} \right\rangle,\, \left\langle e_2/\rho_2 c^2 \right\rangle,\, \left\langle \rho_2/\rho_1 \right\rangle,\, \left\langle p_{b 2}/p_2 \right\rangle,\, \left\langle B_{\perp 2}/B_{\perp 1} \right\rangle$ as functions of relative Lorentz factor $\left\langle\gamma_{21}\right\rangle$, for given tangential magnetization $\left\langle\sigma_\perp\right\rangle$ and anisotropy parameter $\xi$ of a random magnetic field.
    The grey dashed lines represent solutions given by quasi-normal approximation, directly adopted from Fig.~\ref{fig:1}.
    The colored solid lines indicate exact solutions to jump conditions, with different anisotropy parameters $\xi$ denoted by different colors.
    The exact solutions with different anisotropy but the same $\left\langle\sigma_\perp\right\rangle$ nearly overlap with the same approximated solution, as labeled by the value of $\left\langle\sigma_\perp\right\rangle$.
    The parameters $\left\langle e_2/\rho_2 c^2 \right\rangle$, $\left\langle \rho_2/\rho_1 \right\rangle$ and $\left\langle B_{\perp 2}/B_{\perp 1} \right\rangle$ are normalized with respect to their hydrodynamic values.
	}
    \label{fig:3}
\end{figure*}

For oblique shocks with ordered fields, the jump conditions are determined by three upstream parameters $(\sigma_\perp,\, \theta,\, u_{1s})$.
In principle, the upstream conditions may not be homogeneous,e.g., with fluctuations in density that generate turbulence \citep{mizunoMagnetic2014M, dengMagnetized2017A}.
Here for simplicity, we assume a homogeneous upstream with a uniform velocity $u_{1s}$ along the shock normal, but permeated with a random magnetic field.
The distribution of tangential magnetization follows $\sigma_\perp = \sigma_\perp (\theta)$, which can be derived from the function~\eqref{eq:bfieldfunc}.
However, the downstream flow solved from the jump conditions is not strictly homogeneous due to the upstream random field.
For instance, the downstream velocity $u_{2s}$ and density $\rho_2$ are dispersed around their weighted average values $\left\langle u_{2s} \right\rangle$ and $\left\langle \rho_2 \right\rangle$, respectively.
Through jump conditions, the downstream weighted average quantities are fully determined by the weighted average of the tangential magnetization $\left\langle \sigma_\perp \right\rangle$, the magnetic anisotropy parameter $\xi$ and the upstream velocity with respect to the shock $u_{1s}$.

To illustrate the variations of averaged shock parameters analogous to Fig.~\ref{fig:1}, we plot the averaged values of the six parameters $\left\langle u_{2s}\right\rangle,\, \left\langle \gamma_{1s}/\gamma_{21} \right\rangle,\, \left\langle e_2/\rho_2 c^2 \right\rangle,\, \left\langle \rho_2/\rho_1 \right\rangle,\, \left\langle p_{b 2}/p_2 \right\rangle,\, \left\langle B_{\perp 2}/B_{\perp 1} \right\rangle$ in Fig.~\ref{fig:3}.
We take an isotropic random field ($\xi = 1$) and two extremely anisotropic cases ($\xi = 0.1,10$, labeled by different colors) to demonstrate that the jump conditions of shocks with random magnetic fields are insensitive to the field anisotropy.
Since the quasi-normal approximation developed in section~\ref{sec:quasinormal} only depends on two parameters $\sigma_\perp$ and $\gamma_{21}$, here we attempt to estimate these six parameters by directly taking the approximated results from Fig.~\ref{fig:1} as references (grey dashed lines).
Despite small deviations, the quasi-normal approximation generally fits well with the averaged shock parameters if we take $\left\langle \sigma_\perp \right\rangle$ as an effective tangential magnetization and $\left\langle \gamma_{21} \right\rangle$ as an effective relative Lorentz factor.
Furthermore, the asymptotic analytical solution in the ultra-relativistic limit is still applicable to shocks with random fields.
Therefore, the jump conditions of relativistic shocks with random magnetic fields are governed by the tangential magnetization as well, although this tangential magnetization $\left\langle \sigma_\perp \right\rangle$ is an average value weighted averaged over all solid angles.

The most obvious deviation of the exact solutions from our approximation in Fig.~\ref{fig:3} is the rising yellow curve for $\left\langle \sigma_\perp \right\rangle = 0.01,\, \xi = 10$ at $\left\langle \gamma_{21} \right\rangle \lesssim 10$ in the last panel.
This can be explained by the last panel of Fig.~\ref{fig:1}, where we show the amplification of tangential field component deviate from our approximation for weakly magnetized quasi-parallel shocks ($\sigma_\perp = 0.01,\, \theta = 10^\circ$) in the near-Newtonian regime.
The $\xi = 10$ case in Fig.~\ref{fig:3} also represents a quasi-parallel shock where field lines are nearly directed along the shock normal, but with a highly anisotropic random field instead of a globally ordered field.
We thus expect the $\left\langle \sigma_\perp \right\rangle = 0.01,\, \xi = 10$ solution in the last panel of Fig.~\ref{fig:3} would also reach a peak and then decrease as $\left\langle \gamma_{21} \right\rangle \rightarrow 1$, similar to the quasi-parallel solution shown in the last panel of Fig.~\ref{fig:1}.
However, the near-Newtonian solutions are not shown here because they are not fast magnetoacoustic shocks, which is discussed in the last paragraph of this section.

\begin{figure}
	\includegraphics[width=\columnwidth]{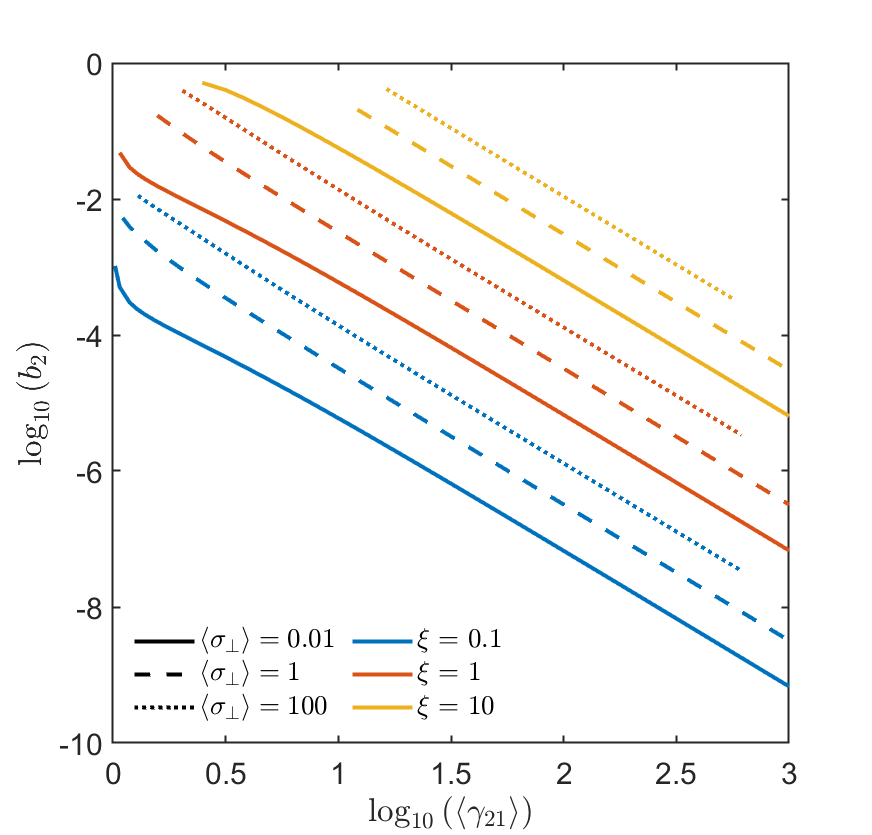}
    \caption{
    Downstream anisotropy parameters $b_2 \equiv {2\left\langle B_{\parallel 2}^2 \right\rangle}/{\left\langle B_{\perp 2}^2 \right\rangle}$ of shock-compressed random magnetic field, as functions of relative Lorentz factor $\left\langle\gamma_{21}\right\rangle$, for given tangential magnetization $\left\langle\sigma_\perp\right\rangle$ and upstream anisotropy parameter $\xi$.
    The solid, dashed and dotted lines denote the cases for $\left\langle\sigma_\perp\right\rangle = 0.01,1,100$, respectively, with different upstream anisotropy parameters $\xi$ denoted by different colors.
    }
    \label{fig:4}
\end{figure}

The anisotropy parameter $b_2 \equiv {2\left\langle B_{\parallel 2}^2 \right\rangle}/{\left\langle B_{\perp 2}^2 \right\rangle}$ of the shock-compressed random field in the downstream is shown in Fig.~\ref{fig:4}.
Different colors represent the cases for different upstream anisotropy $\xi$, while the solid, dashed and dotted lines indicate $\left\langle\sigma_\perp\right\rangle = 0.01,1,100$, respectively.
It is obvious that the downstream anisotropy scales as $b_2 \propto \left\langle \gamma_{21} \right\rangle^{-2}$ because the tangential field component is amplified by a factor of $4\left\langle \gamma_{21} \right\rangle$ while the parallel field components are approximately the same across the shock.
Since the upstream anisotropy parameters are related by $b_1 = \xi^2$, we find that the post-shock anisotropy parameter $b_2$ would always be smaller than the pre-shock anisotropy parameter $b_1$ if the shock is only weakly magnetized ($\left\langle\sigma_\perp\right\rangle \ll 1$).
However, a strongly magnetized shock ($\left\langle\sigma_\perp\right\rangle \gtrsim 1$) boosts the post-shock anisotropy parameter, i.e., making the downstream magnetic field more isotropic.
It is nevertheless a general trend that a higher upstream anisotropy parameter $b_1$ directly leads to a higher downstream anisotropy parameter $b_2$, even with the boost from a strongly magnetized shock.

We highlight here that we only consider the shock-compressed field in this study.
The amplification and generation of magnetic field from other effects (e.g., Weibel instability, \citet{medvedevGeneration1999A}; turbulence, \citet{mizunoMagnetic2014M, dengMagnetized2017A}) are not taken into account.
The Weibel instability would generate a random field nearly confined to the shock plane \citep{medvedevGeneration1999A}, which is believed to have an anisotropy parameter $b_2 \ll 1$ \citep{gillConstraining2020M, teboulImpact2021M} and operates at low magnetization ($\sigma \lesssim 10^{-3}$) \citep{sironiParticle2009A, sironiParticle2011A}.
Other effects such as turbulence and drift motion may isotropize the field and increase the anisotropy parameter $b_2$.

It is obvious in Fig.~\ref{fig:3} and Fig.~\ref{fig:4} that the curves near $\left\langle \gamma_{21} \right\rangle \sim 1$ are not shown.
This is because we are only interested in fast magnetoacoustic shocks believed to be efficient particle accelerators \citep{1999JPhG...25R.163K, sironiParticle2009A}, but the near-Newtonian quasi-parallel shocks are not fast shocks \citep[see, e.g., Figure 1 in][]{sironiRelativistic2015M}
\footnote{To see this point, consider a strictly parallel shock.
The jump conditions are given by hydrodynamic solutions \citep{blandfordFluid1976PF}.
For near-Newtonian shocks, we have $\gamma_{1s} \simeq \hat{\Gamma}(\gamma_{21}-1)+1$.
A necessary condition for a fast shock to develop is that the shock velocity exceeds fast magnetoacoustic wave speed in the upstream rest frame, i.e., $\gamma_{1s}>\sqrt{1+\sigma}$ for a cold upstream.
Then we get a lower limit of the relative Lorentz factor $\gamma_{21} > 1+\sigma/2\hat{\Gamma}$ for the near Newtonian shock ($\gamma_{21}\sim 1$) to be a fast shock.
Consequently, any near-Newtonian parallel shock that does not satisfy this condition is not a fast shock.
}.
Since we take the weighted average over all possible solid angles for random fields, inside the integral there would be an interval near $\theta \sim 0$ where there is no fast shock solution.
The solutions that are not fast shocks are thus omitted in Fig.~\ref{fig:3} and Fig.~\ref{fig:4}.

\section{Conclusions and discussion}
\label{sec:conclusions}

In this paper, we revisit the jump conditions of relativistic oblique shocks with ordered or random magnetic fields.

For relativistic shocks permeated by ordered fields, we explore the jump conditions with an arbitrary upstream magnetization parameter $\sigma$ and an arbitrary obliquity angle $\theta$.
The tangential degree of magnetization $\sigma_\perp \equiv \sigma \sin^2 \theta = B_\perp^2/\rho h$ instead of the total magnetization $\sigma$ is identified as the characteristic parameter of the jump conditions.
When the tangential field component is at least comparable to the parallel component ($\theta \gtrsim 30^\circ$) and the shock is at least mildly relativistic (relative Lorentz factor $\gamma_{21} \gtrsim 3$ between upstream and downstream), the jump conditions are approximately determined by tangential magnetization $\sigma_\perp$, insensitive to the magnetic obliquity.
In this case, the approximation holds because the downstream velocity is nearly directed along the shock normal (i.e., quasi-normal approximation), such that the parallel component of the magnetic field is decoupled from the jump conditions.

For relativistic shocks with a random magnetic field in the upstream, the downstream flow is no longer homogeneous but can be described by the averaged values $\left\langle Q \right\rangle$ of flow parameters $Q$.
With $\left\langle \sigma_\perp \right\rangle$ as an effective tangential magnetization and $\left\langle \gamma_{21} \right\rangle$ as an effective relative Lorentz factor, the averaged shock parameters are still well-approximated by quasi-normal approximation, insensitive to the anisotropy of the random field.
We therefore conclude that the jump conditions of relativistic MHD shocks are governed by the magnetization of the tangential magnetic field component, regardless of whether the field lines are ordered or randomized.
The downstream field lines are compressed and amplified in the tangential direction.
However, the inhomogeneity of the downstream flow induced by the upstream random field alone may generate turbulent motions.
The turbulence and global motions of the fluid could in principle stretch the downstream field lines into a more isotropic configuration \citep{gillConstraining2020M}, which is beyond the scope of this study.

It was shown previously that the appropriately normalized flow parameters are insensitive to relative Lorentz factor $\gamma_{21}$ for relativistic perpendicular shock ($\gamma_{21} \gg 1$) \citep{zhangGamma2005Aa, mimicaInternal2007A}, which follow the analytical solutions first presented in \citet{kennelConfinement1984A}.
Here, we prove that these asymptotic analytical solutions also hold for relativistic oblique shocks ($\gamma_{21} \gtrsim 10$, but with the $\sigma$ parameter replaced by $\sigma_\perp$), regardless of whether field lines are ordered or randomized.
This conclusion could serve as the starting point of theoretical models of shock-powered GRBs or FRBs with a more general field geometry.
The quasi-normal approximation could be applied as a quick check for numerical simulations of particle acceleration \citep[e.g.,][]{sironiParticle2009A} and shock formation \citep[e.g.,][]{mimicaInternal2007A}.
However, the MHD jump conditions only describe the shock-compressed magnetic field and omit the kinetic effects (e.g., turbulence, instability and wave generation), therefore incapable to be fully consistent with results from numerical simulations \citep{plotnikovPerpendicular2018M, bretCan2020A}.
Nevertheless, in the high $\sigma$ regime, the kinetic effects are expected to be suppressed as long as the shock is not quasi-parallel, so that the shock structure can be well-approximated by the MHD jump conditions \citep{sironiParticle2009A, sironiMaximum2013A, bretCan2020A}.

In section~\ref{sec:general}, we discuss two definitions of magnetization parameter and their distinctions for oblique shocks. One is defined as the enthalpy density ratio between magnetic field and matter in the comoving frame, which is equivalent to $\sigma$ given by equation~\eqref{eq:sigma}.
The other is defined as the ratio of the Poynting flux to the matter energy flux in the lab frame (or the shock frame in our formalism), which is equivalent to $\sigma_\perp$ given by equation~\eqref{eq:sigma_perp}.
These two definitions are identical for perpendicular shocks, but they are no longer equivalent for oblique shocks.
They can be easily confused with each other, and therefore the distinction is extremely important.
For instance, the jump conditions in \citet{applShock1988A} were derived using the flux ratio definition of magnetization, which only involves the tangential field component.
The method of \citet{applShock1988A} was later adopted in \citet{sironiParticle2009A} to compare the simulations with MHD jump conditions, but the magnetization in \citet{sironiParticle2009A} was defined by the total magnetic field strength.
In this work, we find that the jump conditions of relativistic shocks are governed by $\sigma_\perp$ instead of $\sigma$, which suggests that the definition with flux ratio could be a better choice for magnetization parameter.

In principle, the analysis presented in this work can be naturally extended to a hybrid case, where the magnetic field is composed of both ordered and random components.
We expect that the average tangential magnetization would still be the main ingredient that determines the jump conditions of relativistic shocks.
This general field configuration has been introduced to study the polarization degree and polarization angle in GRB afterglows \citep{granotLinear2003A, teboulImpact2021M}.
It would be interesting to investigate the effect of a hybrid field on MHD shock structure and particle acceleration with numerical simulations.

Here, we only focus on the structure of a single shock.
However, when a relativistic ejecta collides into a medium, the shock-powered GRB/FRB theory involves a two-shock system: a reverse shock propagating into the ejecta and a forward shock propagating into the medium.
The forming condition of the reverse shock or forward shock has been extensively discussed in the form of the Riemann problem, both for hydrodynamic shocks \citep[][and references therein]{2013rehy.book.....R} and perpendicular MHD shocks \citep{zhangGamma2005Aa, 2005JFM...544..323R, gianniosexistence2008A, 2009ApJ...690L..47M, mimicaDeceleration2009A, 2010MNRAS.401..525M, aimechanical2021M}.
Although the Riemann problem for oblique shocks has been solved, the solution generally consists more than two non-linear waves for oblique magnetic fields \citep{1999MNRAS.303..343K, 2006JFM...562..223G}, which makes it notoriously difficult to determine the critical condition for each non-linear wave to develop into a shock.
In the presence of magnetic obliquity, the general forming condition of the reverse shock still merits further investigation.

\section*{Acknowledgements}

We thank the anonymous referee for helpful comments.
J.-Z. Ma acknowledges the support from Tsien Excellence in Engineering Program (TEEP) in Tsinghua University.


\section*{Data Availability}

The code developed to perform the calculation in this paper is available upon request.




\bibliographystyle{mnras}
\bibliography{Shock} 




\appendix

\section{Methods to solve the jump conditions of oblique shocks}
\label{sec:appendix_oblique}

\subsection{Derivation of the governing equations for oblique shocks}

To simplify equations~\eqref{eq:general_jc}, we further define a vector $X^\mu \equiv W^\mu - (W^\beta l_\beta - V^\alpha V_\alpha)l^\mu$ with $u_l \equiv u^\mu l_\mu$, $b_l \equiv b^\mu l_\mu$, $V^\mu \equiv b^\mu u_l - b_l u^\mu$, $W^\mu \equiv (\rho h+b^2)u_l u^\mu + (p+b^2/2)l^\mu - b_l b^\mu$.
We find five scalar equations \citep{1967rhm..book.....L}
\begin{align}
    & \left[\left[\rho u_l\right]\right] = 0\, ,
    \label{eq:scalar1}\\
    & \left[\left[\rho h b_l u_l\right]\right] = \left[\left[W^\alpha V_\alpha\right]\right] = 0\, ,
    \label{eq:scalar2}\\
    & \left[\left[b^2 u_l^2 - b_l^2\right]\right] = \left[\left[V^\alpha V_\alpha\right]\right] = 0\, ,
    \label{eq:scalar3}\\
    & \left[\left[\rho h u_l^2 + p + b^2/2\right]\right] = \left[\left[W^\beta l_\beta - V^\alpha V_\alpha\right]\right] = 0\, ,
    \label{eq:scalar4}\\
    \begin{split}
        & \left[\left[\rho^2 h^2 u_l^2 (1+u_l^2) - 2p(b^2 u_l^2 - b_l^2) + 2\rho h b^2 u_l^2\right]\right] \\
        & = \left[\left[- X^\mu X_\mu - 2 V^\nu V_\nu (W^\beta l_\beta - V^\alpha V_\alpha)\right]\right] = 0\, .
    \end{split}
    \label{eq:scalar5}
\end{align}
We then introduce the following dimensionless quantities 
\citep[see also][for a different treatment]{1987PhFl...30.3045M, 1999JPhG...25R.163K}
\begin{align}
    & \sigma \equiv b^2/\rho h\, ,\\
    & \sigma_l \equiv b_l^2/\rho h\, ,\\
    & A \equiv \sigma_l/u_l\, ,\\
    & C \equiv \sigma u_l - \sigma_l/u_l\, ,\\
    & D \equiv \sigma u_l = A + C\, ,\\
    & H \equiv h/c^2\, ,\\
    & P \equiv p/\rho h = \left(1-1/H\right)\left(1-1/\hat{\Gamma}\right)\, .
\end{align}
With equation (\ref{eq:scalar1}), equations (\ref{eq:scalar2}), (\ref{eq:scalar3}), (\ref{eq:scalar4}), (\ref{eq:scalar5}) can be rewritten as \begin{align}
    & \left[\left[H^3 A\right]\right] = 0\, ,
    \label{eq:a}\\
    & \left[\left[H C\right]\right] = 0\, ,
    \label{eq:c}\\
    & \left[\left[H \left(u_l + P/u_l + D/2u_l^2\right)\right]\right] = 0\, ,
    \label{eq:q}\\
    & \left[\left[H^2\left(1 + u_l^2 - 2PC/u_l + 2D/u_l\right)\right]\right] = 0\, .
    \label{eq:s}
\end{align}
It is thus convenient to introduce invariants $a = H_1^3 A_1,\ c = H_1 C_1,\ q = H \left(u_{l1} + P_1/u_{l1} + D_1/2u_{l1}^2\right),\ s = H_1^2\left(1 + u_{l1}^2 - 2P_1 C_1/u_{l1} + 2D_1/u_{l1}\right)$ which remain the same across the shock front.
When the upstream quantities $(u_{l1}, H_1, \hat{\Gamma}_1, \sigma_1, \sigma_{l1})$ are specified, the invariants $(a,c,q,s)$ are then known and the downstream quantities $(u_{l2}, H_2, \sigma_2, \sigma_{l2})$ can be solved from equations (\ref{eq:a}), (\ref{eq:c}), (\ref{eq:q}), (\ref{eq:s}) if $\hat{\Gamma}_2$ is given.

Let $L \equiv s + H^2(P-1)$ and $J \equiv 3/2 - 2PC/D = 3/2 - 2PH^2/(H^2+a/c)$.
First, notice that equations (\ref{eq:q}) and (\ref{eq:s}) can be combined to give
\begin{equation}
    q H_2 u_{l2} + H_2^2 J_2 D_2/u_{l2} - L_2 = 0\, .
    \label{eq:hyperbola}
\end{equation}
Equations (\ref{eq:s}) and (\ref{eq:hyperbola}) can be further manipulated into a linear equation for $u_{l2}$ which yields
\begin{equation}
    u_{l2} = \frac{q L_2/2 + q H_2^2 P_2 J_2 - H_2^3 J_2^2 D_2}{H_2\left(q^2/2 + q^2 J_2 - L_2 J_2\right)}\, .
\end{equation}
Substituting this expression into equation (\ref{eq:hyperbola}) gives a polynomial for $H_2$
\begin{equation}
    \begin{split}
        & \left(q L_2/2 + q H_2^2 P_2 J_2 - H_2^3 J_2^2 D_2\right)\left(q^2 s - q^2 H_2^2 + q H_2^3 J_2 D_2 - L_2^2\right) \\
        & = H_2^3 D_2\left(q^2/2 + q^2 J_2 - L_2 J_2\right)^2\, ,
    \end{split}
\end{equation}
which can only be solved numerically for the general cases.

\subsection{Shock-forming conditions of fast-mode shocks}

In the fluid rest frame, MHD wave admits three wave modes, i.e., fast mode with dimensionless speed $\beta_{\mathrm{fast}}$, slow mode with dimensionless speed $\beta_{\mathrm{slow}}$ and Alfven mode with dimensionless speed $\beta_{\mathrm{A}}$.
Here, $\beta_{\mathrm{A}}$ is the dimensionless Alfven wave propagation speed along the shock normal given by \citep{1999JPhG...25R.163K, 1999MNRAS.303..343K}
\begin{equation}
    \beta_{\mathrm{A}} = \sqrt{\frac{\sigma}{1+\sigma}} \cos\theta\, ,
\end{equation}
where $\theta$ is the angle between magnetic field lines and shock normal measured in fluid rest frame.
The dimensionless fast/slow magnetoacoustic wave speeds are determined by an equation for $\beta$
\begin{equation}
   \beta^4 - \left[\beta_\mathrm{so}^2 + (1-\beta_\mathrm{so}^2)\sigma/(1+\sigma) + \beta_\mathrm{so}^2\beta_\mathrm{A}^2\right]\beta^2 + \beta_\mathrm{so}^2\beta_\mathrm{A}^2 = 0\, ,
   \label{eq:fastspeed}
\end{equation}
where $\beta_\mathrm{so} \equiv \sqrt{\hat{\Gamma}p/\rho h} = \sqrt{\hat{\Gamma} P}$ is the dimensionless sound speed.
Generally, we have $\beta_{\mathrm{fast}} \ge \beta_{\mathrm{A}} \ge \beta_{\mathrm{slow}}$, and we are only interested in the fast magnetoacoustic shocks which are believed to be efficient particle accelerators \citep{1999JPhG...25R.163K, sironiParticle2009A}.

The dimensionless speed of shock propagation measured in the fluid rest frame is defined as \citep{1987PhFl...30.3045M}
\begin{equation}
    \beta_{\mathrm{sh}} = u_l/\sqrt{1+u_l^2}\ .
\end{equation}
Let
\begin{equation}
    \begin{split}
        \alpha & \equiv h/\rho - (b_l^2/u_l^2 - b^2)/\rho^2 \\
     & = h(1+\sigma-\sigma_l/u_l^2)/\rho
    = H c^2(u_l + C)/\rho u_l\, .
    \end{split}
\end{equation}
A fast-mode shock can be formed if and only if the following conditions are satisfied: (a) $\beta_{\mathrm{sh},1} > \beta_{\mathrm{fast},1} > \beta_{\mathrm{A},1}$, (b) $\beta_{\mathrm{fast},2} > \beta_{\mathrm{sh},2} > \beta_{\mathrm{A},2}$, (c) $\alpha_1 > \alpha_2 > 0$ \citep{1967rhm..book.....L, 1987PhFl...30.3045M}.
Exploiting jump conditions (\ref{eq:scalar1}) and (\ref{eq:c}), we find that condition (c) is equivalent to two conditions, i.e. (c1) $\left[\left[H u_l\right]\right] > 0$ and (c2) $u_{l2} + C_2 > 0$.

In conclusion, there are four conditions to check for fast-mode MHD shocks: 
\begin{align}
    & \beta_{\mathrm{sh},1} > \beta_{\mathrm{fast},1}\, ,
    \label{eq:form1}\\
    & \beta_{\mathrm{fast},2} > \beta_{\mathrm{sh},2} > \beta_{\mathrm{A},2}\, ,
    \label{eq:form2}\\
    & \left[\left[H u_l\right]\right] > 0\, ,
    \label{eq:form3}\\
    & u_{l2} + C_2 > 0\, .
    \label{eq:form4}
\end{align}

Under the cold upstream assumption ($e_1 = p_1 = 0$), in upstream region the dimensionless specific enthalpy, pressure and sound speed are $H_1 = 1$, $P_1 = 0$, $\beta_{\mathrm{so},1} = 0$, respectively.
According to equation (\ref{eq:fastspeed}), this directly gives the fast wave speed in region 1 as $\beta_{\mathrm{fast},1} = \sqrt{\sigma_1/(1+\sigma_1)}$, Consequently, shock-forming condition (\ref{eq:form1}) suggests that $\sigma_1$ needs to be smaller than $u_{1s}^2$ in order for a fast-mode shock to develop.

\section{Derivation of the solution for quasi-normal approximation}
\label{sec:appendix_quasinormal}

Assuming a cold upstream flow where $e_1=p_1=0$, equations (\ref{eq:3}) and (\ref{eq:4}) can be expressed as
\begin{align}
    &\gamma_{1s}c^2 \left[1+(1-r)\sigma_\perp\right]=\gamma_{2s}h_2\, ,
     \label{eq:3p}\\
    &u_{1s}c^2 \left[1+(1-r^2)\frac{\sigma_\perp}{2\beta_{1s}^2}\right]= u_{2s}h+\frac{p_2}{\rho_2 u_{2s}}\, .
     \label{eq:4p}
\end{align}
Multiplying equation (\ref{eq:3p}) by $\gamma_{2s}$ and equation (\ref{eq:4p}) by $u_{2s}$, we find
\begin{align}
    &\frac{\gamma_{21}}{1-\beta_{1s}\beta_{2s}} c^2 \left[1+(1-r)\sigma_\perp\right]=\gamma_{2s}^2 h_2\, ,\\
    &\frac{\gamma_{21}\beta_{1s}\beta_{2s}}{1-\beta_{1s}\beta_{2s}}c^2 \left[1+(1-r^2)\frac{\sigma_\perp}{2\beta_{1s}^2}\right]= u_{2s}^2 h_2+\frac{p_2}{\rho_2}\, .
\end{align}
Subtracting two equations above gives
\begin{equation}
    \frac{e_2}{\rho_2 c^2}=\gamma_{21}-1-\frac{\gamma_{21}\beta_{21}^2(1-\beta_{2s}^2)}{2\beta_{2s}(\beta_{2s}+\beta_{21})}\sigma_\perp\, .
\end{equation}
Combining this equation with equation (\ref{eq:3p}) by exploiting the equation of state $e_2=p_2/(\hat{\Gamma}-1)$, we find
\begin{equation}
\begin{split}
    &\gamma_{21}(1+\beta_{21}\beta_{2s})\left[1+(1-r)\sigma_\perp\right]\\
    &=1+\hat{\Gamma}\left[\gamma_{21}-1-\frac{\gamma_{21}\beta_{21}^2(1-\beta_{2s}^2)}{2\beta_{2s}(\beta_{2s}+\beta_{21})}\sigma_\perp\right]\, ,
\end{split}
\end{equation}
which by multiplying with $\beta_{2s}(\beta_{2s}+\beta_{21})/\gamma_{21}\beta_{21}$ yields
\begin{equation}
\begin{split}
    &\beta_{2s}(\beta_{2s}+\beta_{21})\left[\beta_{2s}-(\hat{\Gamma}-1)\frac{\beta_{21}}{1+\sqrt{1-\beta_{21}^2}}\right]\\
    &=\left[\beta_{2s}+\left(1-\frac{\hat{\Gamma}}{2}\right)\beta_{21}\right](1-\beta_{2s}^2)\sigma_\perp\, .
\end{split}
\end{equation}

\section{Solutions for contact discontinuities}
\label{sec:appendix_cd}

Contact discontinuities can be treated as special cases of discontinuity surfaces where there is no mass flux across the surface, i.e., $\rho_1 u_{l1} = \rho_2 u_{l2} = 0$.
This condition directly yields $u_{l1} = u_{l2} = 0$, i.e., the velocities normal to the discontinuity surface  are the same on either side.
The other jump conditions $\left[\left[V^\mu\right]\right] = \left[\left[W^\mu\right]\right] = 0$ would then yield $\left[\left[b_l\right]\right] = \left[\left[p + b^2/2\right]\right] = 0$.
More rigorously, if $b_{l1} = b_{l2} \neq 0$, we get $\left[\left[p\right]\right] = \left[\left[b^\mu\right]\right] = \left[\left[u^\mu\right]\right]= 0$.

We thus conclude that an oblique contact discontinuity ($b_{l1} = b_{l2} \neq 0$) is fundamentally different from a rigorous $90^\circ$ contact discontinuity ($b_{l1} = b_{l2} = 0$).
The pressure, velocity and magnetic field are continuous across an oblique contact discontinuity, whereas a $90^\circ$ contact discontinuity may have different tangential velocities (as in the hydrodynamic case) and different tangential magnetic fields on either side of the front.
However, there is no rigorous $90^\circ$ contact discontinuity in reality.
Therefore for astrophysical applications, the pressure, velocity and magnetic field would always be continuous across any contact discontinuity, whereas jumps in density and specific enthalpy are permitted.
This would invalid the assumption made in \citet{zhangGamma2005Aa} that on one side of the contact discontinuity the fluid is magnetized whilst on the other side the fluid is strictly hydrodynamical.

\bsp	
\label{lastpage}
\end{document}